\documentclass[letterpaper]{article}
\usepackage{gen2ic_preprint}
\usepackage[hyphens]{url}
\usepackage{graphicx}
\usepackage{natbib}
\usepackage{caption}
\usepackage{algorithm}
\usepackage{algorithmic}
\usepackage{amsmath}
\usepackage{booktabs}
\usepackage{multirow}
\usepackage{colortbl}
\usepackage[most]{tcolorbox}
\definecolor{tablegray}{gray}{0.92}
\newtcolorbox[auto counter, number within=section]{promptbox}[2][]{title={Prompt Box~\thetcbcounter: #2},
  colback=black!2,
  colframe=black!20,
  boxrule=0.4pt,
  arc=2pt,
  left=6pt,
  right=6pt,
  top=6pt,
  bottom=6pt,
  #1
}

\title{Gen2-IC: Bridging Generative Models and Image Codecs through Latent Transport}
\author{
Yinhuan Huang \quad Hao Cao \quad Pu Chen\\
Wenqi Guo \quad Jungong Han \quad Zhijin Qin\textsuperscript{*}
}
\affiliations{
Tsinghua University\\
huangyh24@mails.tsinghua.edu.cn \quad qinzhijin@tsinghua.edu.cn\\
\textsuperscript{*}Corresponding author.
}
\begin{document}

\maketitle

\begin{abstract}
Diffusion-based image compression has achieved strong perceptual quality at ultra-low bitrates. 
However, existing codecs are often tied to specific backbones and specialized components, making diverse, rapidly evolving generative models difficult to reuse.
This raises a natural question: \emph{Can modern generative foundation models be connected to image compression through a simple and extensible interface?} 
Two insights guide our design: stronger generative priors make a simpler codec interface viable, and generation and compression can be intrinsically linked through latent transport. 
We therefore propose Gen2-IC with two stages: 
(1) \textit{Latent Compression} maps clean image latents to entropy-constrained latents; and (2) \textit{Latent Transport} refines them with one near-terminal update based on the pretrained model. Gen2-IC requires neither auxiliary conditioning signals nor task-specific backbone modifications. With lightweight adaptation and no distillation, it supports fast encoding and one-step decoding across multiple bitrates. We validate Gen2-IC on SD-2.1, SANA-1.5, FLUX.1-dev, and Qwen-Image-2512, spanning U-Net and Transformer architectures as well as diffusion and flow-matching formulations. With stronger priors, Gen2-IC delivers gains below 0.05 bpp: the Qwen variant leads diffusion-based generative codecs in reconstruction fidelity (PSNR), perceptual similarity (LPIPS and DISTS), and recognizer-based semantic fidelity (OCR CER/WER and face-ROI similarity) across four benchmarks.
\end{abstract}

\section{Introduction}
Lossy image compression aims to reduce bitrate while preserving reconstruction fidelity and perceptual quality. At ultra-low bitrates, the severe information bottleneck intensifies the perception--distortion trade-off~\cite{Blau_2018_CVPR,perception}. Under this constraint, distortion-oriented objectives (e.g., MSE) tend to average uncertain details, causing blur and texture loss. Perceptual losses (e.g., LPIPS~\cite{lpips} and DISTS~\cite{dists}) and adversarial training~\cite{gan} improve visual realism~\cite{perceptioncodec,hific,ms-illm}, but remain insufficient to preserve semantic fidelity, leading to errors such as corrupted text or facial identity drift.

\begin{figure}[t]
  \centering
  \includegraphics[width=1.0\linewidth]{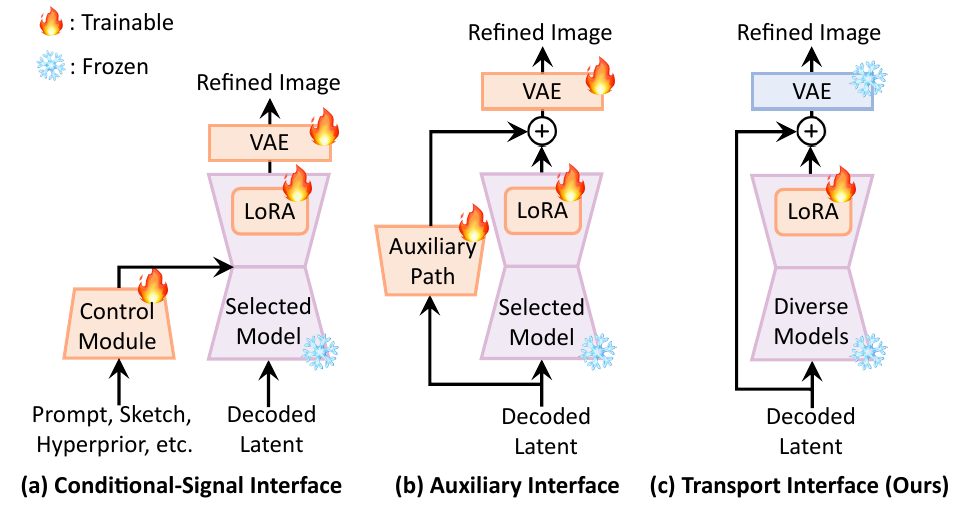}
  \caption{Generative-codec interfaces. Existing methods formulate generative compression as conditional generation, coupling a selected model with conditioning signals or auxiliary paths. Gen2-IC instead formulates reconstruction as latent transport and directly refines the decoded latent through a single path, enabling lightweight adaptation and one-step, multi-rate decoding across diverse models.}
  \label{fig:motivation}
\end{figure}

\begin{figure*}[t]
  \captionsetup{type=figure}
  \includegraphics[width=1.0\textwidth]{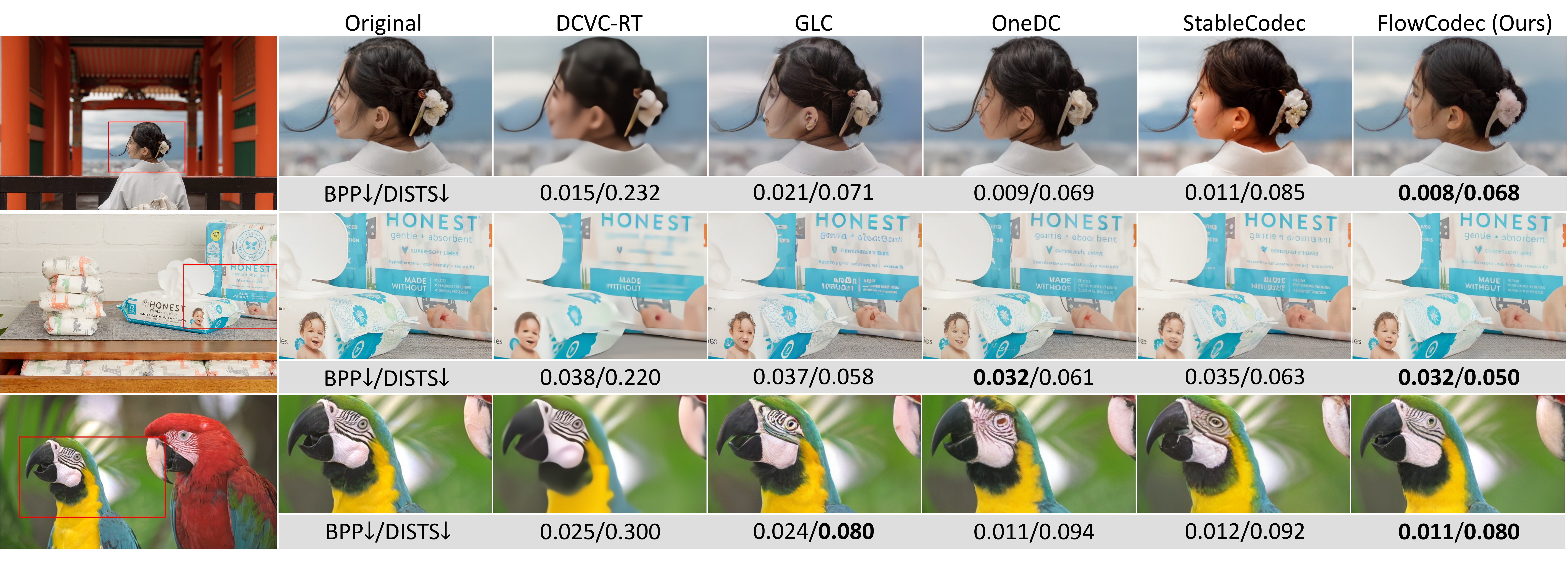}
  \caption{Source fidelity at ultra-low bitrates. Compared with RDVQ~\cite{RDVQ}, OSCAR~\cite{oscar}, StableCodec~\cite{StableCodec}, and DiT-IC~\cite{DiT-IC}, Gen2-IC-Qwen better preserves text and facial details while using the lowest bitrate, with the clearest gains in semantically sensitive regions where small errors alter meaning or identity.}
  \label{fig:Visual_example_0}
\end{figure*}

Modern generative foundation models are rapidly improving their ability to model text, faces, objects, and fine visual structure~\cite{stable-diffusion,sana1.5,flux2024,qwenimagetechnicalreport}. These models provide increasingly powerful generative priors for recovering details omitted by a limited bit budget. However, a core challenge remains: the objectives of generation and compression are not naturally aligned. Generation favors diverse, plausible outputs, whereas compression requires faithful reconstruction of a specific source without semantic drift.

As illustrated in Figure~\ref{fig:motivation}, existing generative codecs bridge this gap by casting reconstruction as conditional generation and coupling a selected generative model with conditioning signals or auxiliary modules~\cite{lei2023text+,perco,StableCodec,oneDC,DiT-IC}. To reduce decoding latency, one-step variants further use distillation objectives or rate--timestep mappings~\cite{StableCodec,oneDC,oscar,DiT-IC}. Although effective, these mechanisms often remain tightly tied to particular backbones and native sampling processes. Consequently, extending these methods to new generative models across diverse architectures and generative processes requires model-specific codec redesign, while adapting larger models raises the costs, leaving their scaling potential underexplored in image compression. This raises a central question: \emph{Can modern generative foundation models be connected to image compression through a simple and extensible interface?}

To answer this, we move beyond model-specific conditioning and revisit the intrinsic relationship between generation and compression. This perspective yields two key insights: (1) by better compensating for information loss at ultra-low bitrates, a stronger decoder-side prior can maintain reconstruction quality with a direct, single-path codec and fewer model-specific components; and (2) generation and compression can be coupled at the mechanism level. 
Specifically, a generative process transports noise toward the data distribution in latent space, whereas compression maps source data to an entropy-constrained latent distribution. This suggests a latent-transport view for integrating generative models into image compression: the prior transports decoded latents back toward the clean latent distribution before VAE synthesis.

These insights lead to \textbf{Gen2-IC}, an effective, efficient, and extensible framework that decouples image coding from generative refinement within a single reconstruction path. Specifically, \textbf{Latent Compression} uses a learned codec to map clean latents to entropy-constrained latents, while \textbf{Latent Transport} refines them with one near-terminal update from a pretrained generative model. 
The two stages are designed and optimized separately, allowing Gen2-IC to benefit from advances in both compression and generation as these fields evolve. 
With lightweight backbone adaptation, it supports fast encoding and distillation-free one-step decoding across multiple bitrates, without additional conditioning or model-specific backbone redesign.

As previewed in Figure~\ref{fig:Visual_example_0}, Gen2-IC preserves text and facial details more faithfully while using the lowest bitrate in each comparison. Extensive experiments further show that the same formulation remains effective across diverse generative models, while stronger priors yield additional perceptual and semantic gains. This perspective offers a practical route for image compression to benefit directly from advances in generative modeling. Our contributions are as follows:
\begin{enumerate}
    \item We formulate the integration of generative models into image compression as latent transport, shifting the focus from codec design for a single generator to a reusable interface across evolving generative-model families.

    \item We introduce Gen2-IC, a decoupled design in which Latent Compression and Latent Transport are optimized separately, with lightweight backbone adaptation. It supports fast encoding, one-step decoding, and multi-rate operation without additional conditioning, prior-specific backbone redesign, or distillation.

    \item We validate the same formulation across SD-2.1, SANA-1.5, FLUX.1-dev, and Qwen-Image-2512. A matched SD-2.1 comparison against OSCAR demonstrates quality and efficiency gains independent of model scale, while the larger FLUX and Qwen variants lead in perceptual quality and recognizer-based text and face preservation.
\end{enumerate}

\begin{figure*}[t]
  \centering
  \includegraphics[width=1.0\textwidth]{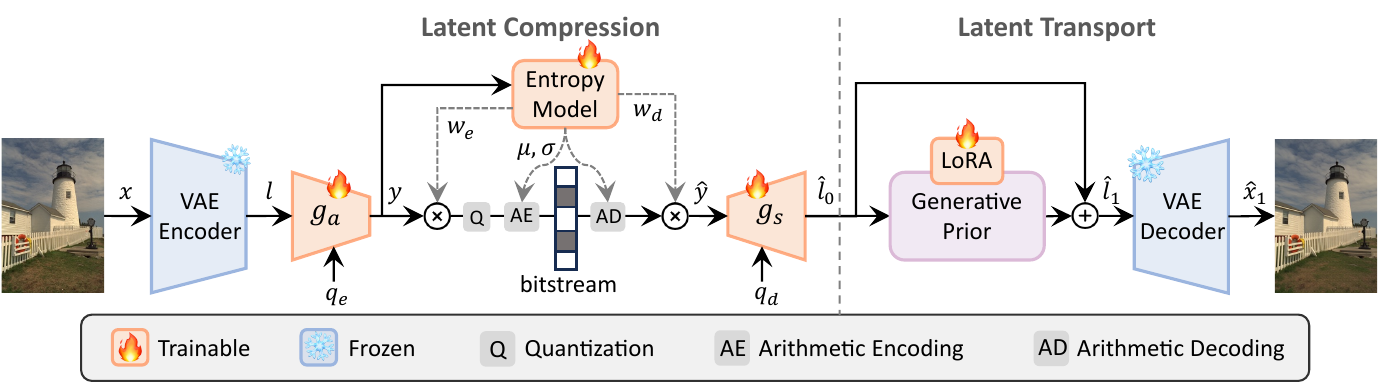}
  \caption{Overview of Gen2-IC. Latent Compression encodes the clean VAE latent $l$ into a bitstream and reconstructs an entropy-constrained latent $\hat{l}_0$. Latent Transport refines $\hat{l}_0$ through one near-terminal update before VAE decoding. The two stages are optimized separately, allowing the same formulation to be instantiated across diverse generative models.}
  \label{fig:framework}
\end{figure*}

\section{Related Work}
\paragraph{Learned and Perceptual Image Compression.}
Learned codecs follow the transform-coding pipeline of traditional codecs~\cite{jpeg,vtm}, replacing hand-crafted components with neural networks for end-to-end nonlinear transforms and entropy estimation, often strengthened by hyperpriors and context models~\cite{lic-balle2018variational,lic-elic,lic-tcm,dcvc-rt}. This paradigm has delivered strong rate--distortion performance and remains computationally practical, but distortion-oriented training often yields blur and texture loss at low rates. Perceptual codecs therefore add learned perceptual losses and adversarial training~\cite{perceptioncodec,hific,ms-illm}, while codebook-based methods exploit discrete representations and nonlinear modeling capacity for perceptual gains at high compression ratios~\cite{vq-gic,vq-DLF,RDVQ}. Nevertheless, at ultra-low bitrates, preserving fine-grained semantics and structure, including facial identity, geometry, and text, remains challenging. Curated data and structure-specific priors can help~\cite{vq-gic,Wang_2021_CVPR,Facecodec}, but add engineering cost.

\paragraph{Generative Priors for Image Compression.}
Diffusion-based codecs exploit generative priors to recover details that rate-constrained representations cannot retain. Existing methods commonly cast reconstruction as conditional generation, guiding a selected generative model with prompts, sketches, or hyperprior features~\cite{lei2023text+,perco,oneDC}. Many rely on iterative or few-step sampling~\cite{cdc,diffeic,rdeic,ResULIC}. Recent methods reduce decoding latency through one-step reconstruction, using distillation or rate- and variance-aware timestep alignment~\cite{StableCodec,oneDC,oscar,DiT-IC}; some additionally introduce auxiliary paths or semantic distillation to improve reconstruction~\cite{StableCodec,oneDC,DiT-IC}. Although effective, these interfaces are usually developed around a particular backbone and sampling process. Meanwhile, generative models are rapidly evolving in both formulation and architecture, from denoising diffusion~\cite{ddpm,stable-diffusion} to flow matching~\cite{lipmanflow} and rectified flow~\cite{liuflow}, and from U-Nets to scalable Transformers~\cite{MMDiT,sana1.5,flux2024,qwenimagetechnicalreport,hunyuanimage,gao2025seedream}. Trained on large-scale image--text corpora, they provide increasingly strong semantic and visual priors. Recent work has also explored diffusion foundation-model pretraining specifically for image compression~\cite{CoD}. Yet the growing diversity of generative models makes model-specific integration difficult to scale. Gen2-IC instead provides a simple and extensible latent-transport interface that connects image compression to existing generative models across architectures and generative processes.

\section{Method}
As shown in Figure~\ref{fig:framework}, Gen2-IC consists of two stages. \textbf{Latent Compression} maps the input image $x$ through the frozen VAE encoder $\mathcal{E}_e(\cdot)$ into a clean latent $l$ and compresses it into an entropy-constrained latent $\hat{l}_0$; \textbf{Latent Transport} then applies one near-terminal update to obtain $\hat{l}_1$, from which the frozen VAE decoder reconstructs $\hat{x}_1=\mathcal{E}_d(\hat{l}_1)$.

\begin{figure*}[t]
  \centering
  \includegraphics[width=1.0\textwidth]{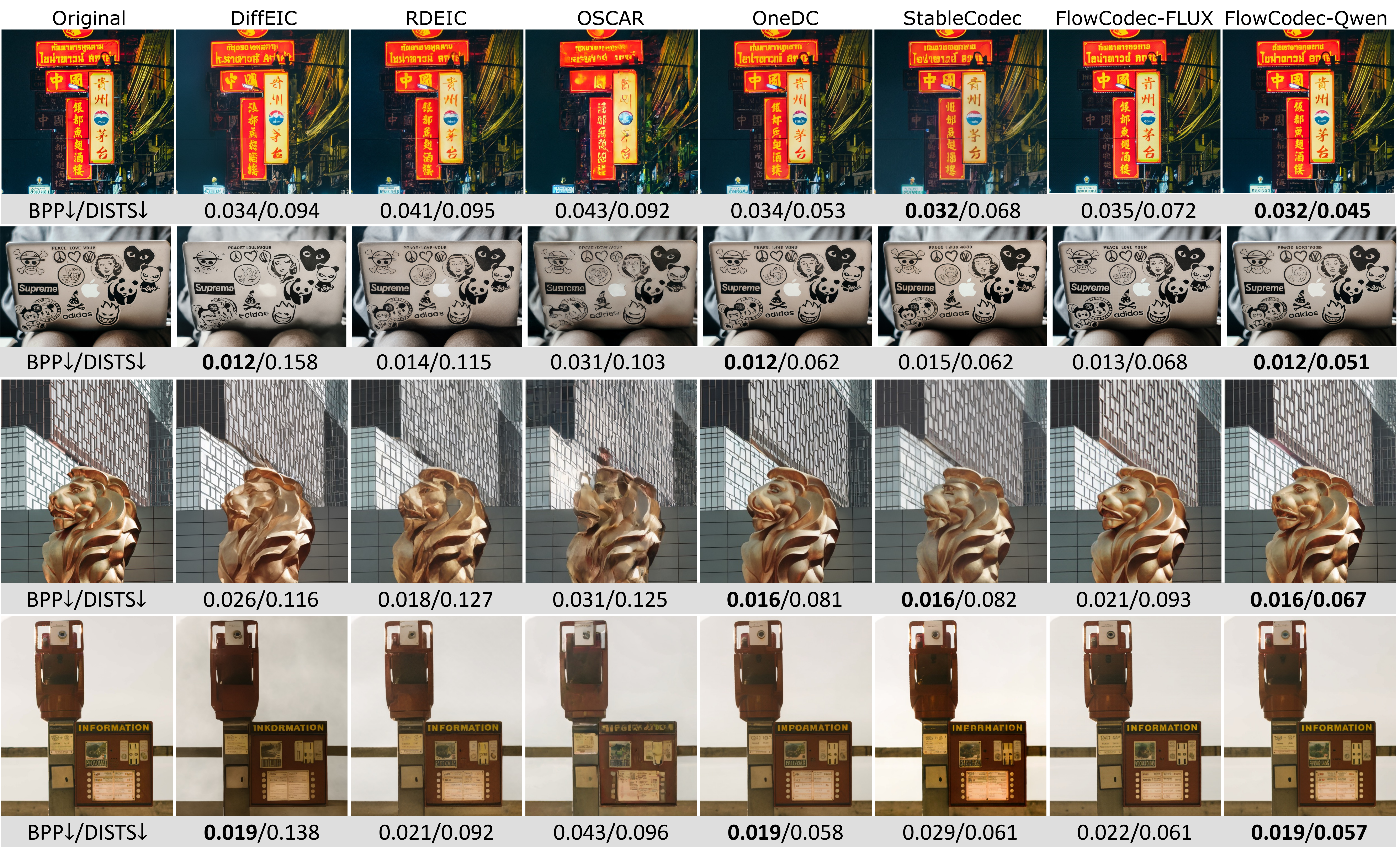}
  \caption{Qualitative comparison at ultra-low bitrates. Cropped regions highlight portraits, text, and fine structures.}
  \label{fig:visual_example_1}
\end{figure*}

\paragraph{Latent Compression.}
Latent Compression learns an efficient multi-rate codec in the VAE latent space. It reconstructs an entropy-constrained latent $\hat{l}_0$ close to the clean latent $l$, providing an image-specific initialization for Latent Transport. To improve parameter efficiency while preserving nonlinear modeling capacity, we adopt depthwise-convolutional transforms~\cite{dcvc-rt}, together with a four-step quadtree autoregressive entropy model $\phi(\cdot)$~\cite{dcvc-dc}. The entropy model predicts Gaussian parameters $(\mu,\sigma)$ for coding the quantized representation $\hat{y}$, together with quantization steps $(w_e,w_d)$. To support multiple bitrates within a single latent-codec model, learnable channel-wise scales $(q_e,q_d)$ modulate the analysis and synthesis transforms $g_a(\cdot)$ and $g_s(\cdot)$~\cite{dcvc-fm}. Combined with $(w_e,w_d)$, these scales control the operating rate. Additional details are provided in the appendix. 
Given an input image $x$, the complete latent-coding process is:
\begin{equation}
\begin{aligned}
\text{Latent extraction:}\;&
l = \mathcal{E}_e(x), \\
\text{Latent encoding:}\;&
y = g_a(l,q_e), \\
\text{Entropy estimation:}\;&
(\mu,\sigma,w_e,w_d)=\phi(y), \\
\text{Quantization:}\;&
\hat{y}
=
\Big(
\mathrm{round}(y\odot w_e-\mu)+\mu
\Big)\odot w_d, \\
\text{Latent decoding:}\;&
\hat{l}_0=g_s(\hat{y},q_d).
\end{aligned}
\label{eq:latent_codec}
\end{equation}

\begin{figure}[t]
  \centering
  \includegraphics[width=1.0\linewidth]{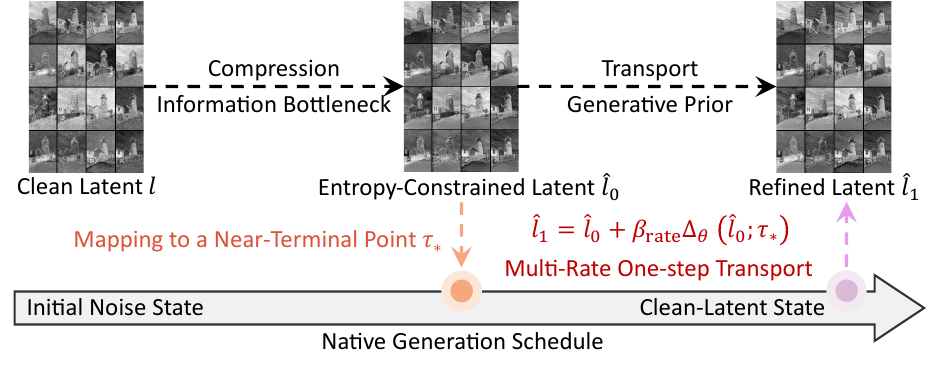}
  \caption{Rate-aware Latent Transport. Gen2-IC maps the entropy-constrained latent $\hat{l}_0$ to a near-terminal state of the generative model's native schedule and applies one update, with $\beta_{\mathrm{rate}}$ controlling the correction strength across bitrates.}
  \label{fig:Latent_transport}
\end{figure}

\paragraph{Latent Transport.}
Latent Compression maps $l$ to an entropy-constrained latent $\hat{l}_0$ through quantization and the information bottleneck. Although $\hat{l}_0$ retains the source content and layout, it may lose fine texture or distort structured regions. Because $\hat{l}_0$ is source-derived and already close to $l$, it provides a more informative initialization than the random noise used in generation. We therefore use $\hat{l}_0$ to construct a state near the end of the generative model's native schedule and apply one scheduler update to obtain $\hat{l}_1$, as illustrated in Figure~\ref{fig:Latent_transport}. The decoded latent itself supplies the image-specific signal, requiring no image-specific prompt or auxiliary semantic input. 
The resulting update is:
\begin{equation}
\hat{l}_1
=
\hat{l}_0
+
\beta_{\mathrm{rate}}\,
\Delta_{\theta}
\bigl(\hat{l}_0;\tau_{\star}\bigr),
\label{eq:near_terminal_update}
\end{equation}
where $\Delta_{\theta}(\hat{l}_0;\tau_{\star})$ denotes the correction induced
by one native scheduler update, and $\tau_{\star}$ is the final update point
in each model's native schedule before VAE decoding. 
Near-terminal placement limits the generative prior's freedom, reducing
over-refinement and hallucination risk.

Unlike approaches that map each bitrate to a pseudo-timestep~\cite{oscar,DiT-IC}, Gen2-IC controls multi-rate refinement
through $\beta_{\mathrm{rate}}$ at the fixed $\tau_{\star}$, using larger
values at lower bitrates and smaller values at higher bitrates. This supports
multiple rates without rate-specific adapters while preserving one-step
decoding.

\begin{figure*}[t]
  \centering
  \includegraphics[width=\textwidth]{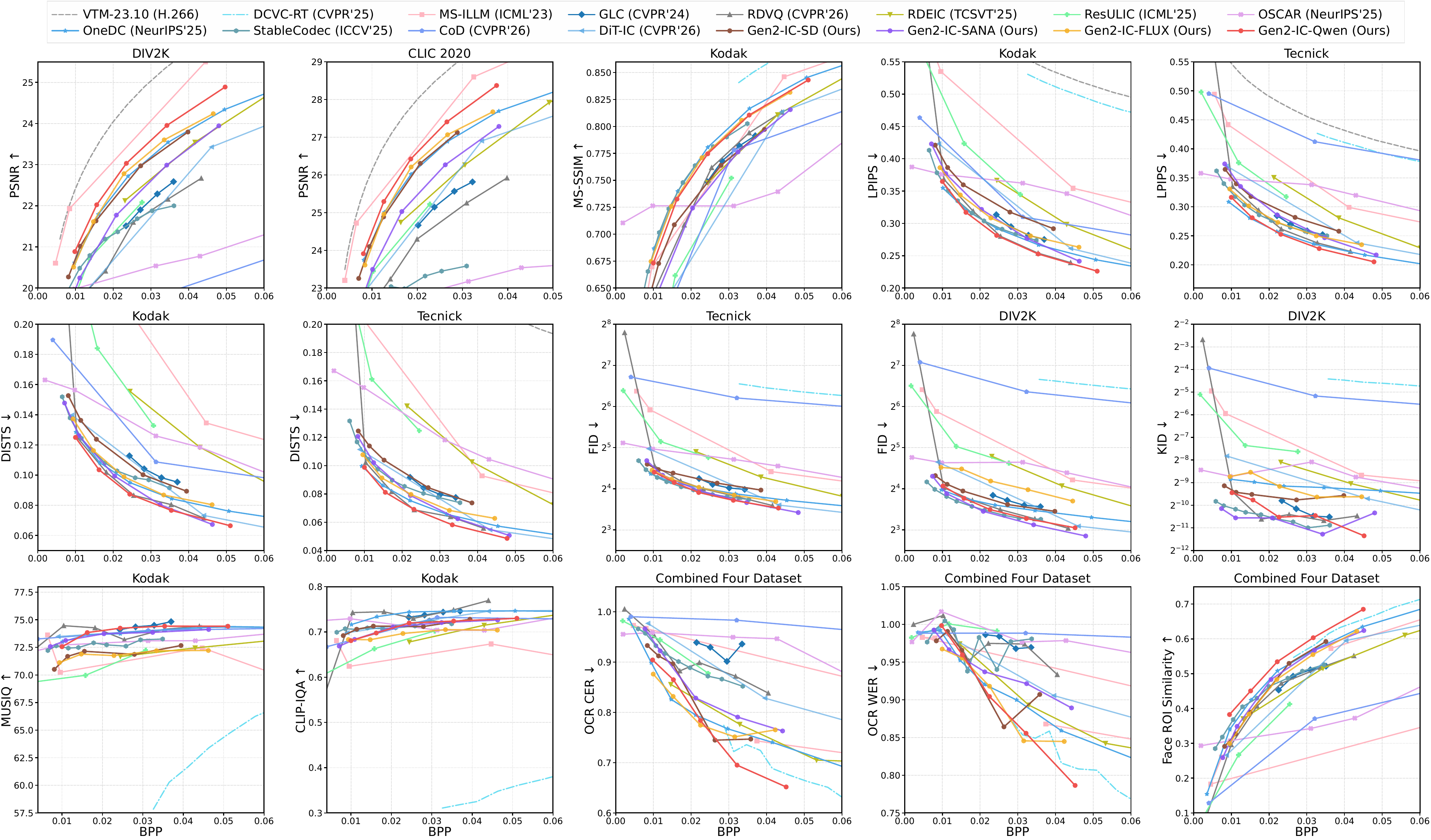}
  \caption{Rate--quality comparison on representative dataset--metric pairs. Higher is better for PSNR, MS-SSIM, MUSIQ, CLIP-IQA, and ROI similarity, while lower is better for LPIPS, DISTS, FID, KID, CER, and WER. CER, WER, and ROI similarity are evaluated over the combined Kodak, Tecnick,
  DIV2K, and CLIC 2020 datasets.}
  \label{fig:rate_quality_curves}
\end{figure*}

\paragraph{Decoupled Optimization.}
Gen2-IC optimizes Latent Compression and Latent Transport in separate stages,
avoiding joint end-to-end training. The codec is trained first and then frozen
while only the LoRA~\cite{lora} parameters are adapted for transport.
The VAE encoder $\mathcal{E}_e(\cdot)$ and decoder $\mathcal{E}_d(\cdot)$ remain frozen throughout. 
The latent codec comprises the analysis transform $g_a(\cdot)$, synthesis transform $g_s(\cdot)$, and entropy model $\phi(\cdot)$. We first optimize these components with a latent-space rate--distortion objective:
\begin{equation}
\mathcal{L}_{\mathrm{LC}}^{(1)}
=
\lambda_{\mathrm{rate}}
\,\mathrm{MSE}\!\left(l,\hat{l}_0\right)
+
R(y),
\end{equation}
where $R(y)$ denotes the bitrate estimated by $\phi(\cdot)$, and $\lambda_{\mathrm{rate}}$ controls the rate--distortion trade-off. This objective bypasses the VAE decoder and accelerates convergence.

\begin{table*}[t]
\centering
\begingroup
\small
\setlength{\tabcolsep}{0.7mm}
\begin{tabular*}{\textwidth}{@{}l@{\hspace{2\tabcolsep}\extracolsep{\fill}}*{13}{r}@{}}
\toprule
& \multicolumn{13}{c}{BD-rate (\%) $\downarrow$} \\
\cmidrule(lr){2-14}
Method
& \multicolumn{3}{c}{Kodak}
& \multicolumn{3}{c}{Tecnick}
& \multicolumn{3}{c}{DIV2K}
& \multicolumn{3}{c}{CLIC 2020}
& \multicolumn{1}{c@{}}{\textbf{Average}} \\
& PSNR & LPIPS & DISTS
& PSNR & LPIPS & DISTS
& PSNR & LPIPS & DISTS
& PSNR & LPIPS & DISTS & \\
\midrule
\rowcolor{tablegray}
\multicolumn{14}{c}{\textit{Distortion-oriented methods}} \\
VTM-23.10 (Intra)
& \underline{-34.11} & 313.84 & 480.92
& \underline{-12.66} & 295.19 & 515.25
& \underline{-18.89} & 285.10 & 543.48
& \underline{-21.32} & 498.64 & 724.83
& 297.52 \\
DCVC-RT (Intra)
& \textbf{-56.39} & 244.62 & 504.80
& \textbf{-46.59} & 237.52 & 586.05
& \textbf{-46.22} & 230.70 & 636.97
& \textbf{-51.33} & 538.81 & 960.16
& 311.59 \\
\midrule
\rowcolor{tablegray}
\multicolumn{14}{c}{\textit{VAE-based methods}} \\
MS-ILLM (anchor)
& 0.00 & 0.00 & 0.00
& 0.00 & 0.00 & 0.00
& 0.00 & 0.00 & 0.00
& 0.00 & 0.00 & 0.00
& 0.00 \\
GLC
& 189.71 & -67.89 & -67.83
& 151.98 & -57.51 & -48.98
& 223.66 & -62.38 & -62.16
& 212.87 & -61.02 & -45.58
& 25.41 \\
RDVQ
& 189.18 & -68.23 & -70.96
& 191.12 & -54.92 & -59.85
& 268.85 & -61.47 & -65.89
& 267.20 & -56.56 & -50.95
& 35.63 \\
\midrule
\rowcolor{tablegray}
\multicolumn{14}{c}{\textit{Diffusion-based methods}} \\
RDEIC
& 72.81 & -52.02 & -39.56
& 82.75 & -31.46 & -3.22
& 108.09 & -35.33 & -4.34
& 120.70 & -19.05 & 13.63
& 17.75 \\
ResULIC
& 138.75 & -41.10 & -38.59
& 108.47 & -41.43 & -32.76
& 195.05 & -41.51 & -20.39
& 165.24 & -35.27 & -12.42
& 28.67 \\
OSCAR
& 426.46 & -39.41 & -49.05
& 489.42 & -5.39 & -4.96
& 602.30 & -14.51 & -20.57
& 839.93 & 21.16 & 36.02
& 190.12 \\
OneDC
& 91.36 & -71.19 & -76.40
& 50.49 & \underline{-66.11} & -64.69
& 94.83 & -67.94 & -70.05
& 75.90 & \underline{-69.79} & -64.72
& \underline{-19.86} \\
StableCodec
& 51.47 & \underline{-74.45} & -78.20
& 117.29 & -65.50 & -65.03
& 212.04 & \underline{-68.08} & -74.75
& 460.96 & -65.90 & -64.13
& 23.81 \\
CoD
& 180.11 & -62.22 & -67.32
& 641.99 & 139.40 & 441.82
& 994.34 & 147.82 & 664.04
& 1666.90 & 339.34 & 1507.01
& 549.43 \\
DiT-IC
& 147.83 & -64.47 & -78.04
& 122.51 & -52.10 & \underline{-66.82}
& 176.14 & -57.11 & -75.98
& 121.72 & -56.79 & \underline{-69.75}
& 3.93 \\
\midrule
\rowcolor{tablegray}
\multicolumn{14}{c}{\textit{Gen2-IC variants}} \\
Gen2-IC-SD
& 71.82 & -62.65 & -72.64
& 67.02 & -52.51 & -53.72
& 102.48 & -57.17 & -66.96
& 75.20 & -52.52 & -51.31
& -12.75 \\
Gen2-IC-SANA
& 114.28 & -69.72 & \underline{-79.58}
& 121.19 & -55.59 & -64.04
& 155.45 & -63.32 & \underline{-77.24}
& 123.80 & -59.28 & -67.48
& -1.79 \\
Gen2-IC-FLUX
& 48.50 & -69.78 & -76.96
& 39.30 & -62.03 & -64.13
& 91.84 & -50.64 & -60.85
& 69.67 & -42.01 & -43.63
& -18.39 \\
Gen2-IC-Qwen
& 44.66 & \textbf{-75.84} & \textbf{-81.21}
& 44.23 & \textbf{-69.25} & \textbf{-71.18}
& 67.41 & \textbf{-72.21} & \textbf{-79.00}
& 46.92 & \textbf{-70.48} & \textbf{-70.66}
& \textbf{-32.22} \\
\bottomrule
\end{tabular*}
\endgroup
\caption{Rate--distortion--perception comparison.
BD-rate (\%; lower is better) relative to MS-ILLM on Kodak, Tecnick, DIV2K,
and CLIC 2020. The Average column aggregates PSNR, LPIPS, and DISTS across all
four datasets.}
\label{tab:bd_rate_ms_illm}
\end{table*}

We then obtain $\hat{x}_0=\mathcal{E}_d(\hat{l}_0)$ and refine the latent codec with an image-space objective:
\begin{equation}
\begin{aligned}
\mathcal{L}_{\mathrm{LC}}^{(2)}
= {} & \lambda_{\mathrm{rate}}
\Big(
\mathcal{L}_1(x,\hat{x}_0)
+
\mathcal{L}_{\mathrm{per}}(x,\hat{x}_0) \\
& \qquad
+
\mathcal{L}_{\mathrm{adv}}(x,\hat{x}_0)
\Big)
+
R(y),
\end{aligned}
\end{equation}
where $\mathcal{L}_1(\cdot)$, $\mathcal{L}_{\mathrm{per}}(\cdot)$~\cite{lpips}, and $\mathcal{L}_{\mathrm{adv}}(\cdot)$~\cite{gan} denote the L1 reconstruction, perceptual, and adversarial losses, respectively. This objective directly improves decoded image $\hat{x}_0$ quality.

After optimizing Latent Compression, we freeze the latent codec and adapt only lightweight LoRA parameters of the pretrained generative backbone. The objective is:
\begin{equation}
\mathcal{L}_{\mathrm{LT}}
=
\mathcal{L}_1(x,\hat{x}_1)
+
\mathcal{L}_{\mathrm{per}}(x,\hat{x}_1)
+
\mathcal{L}_{\mathrm{adv}}(x,\hat{x}_1),
\end{equation}
where $\hat{x}_1=\mathcal{E}_d(\hat{l}_1)$ is decoded from the refined latent $\hat{l}_1$. 
This source-referenced loss drives the LoRA-adapted prior to learn a residual correction at the source-anchored near-terminal state rather than perform unconstrained generation.

\begin{table}[t]
\centering
\begingroup
\small
\setlength{\tabcolsep}{0.7mm}
\begin{tabular*}{\linewidth}{@{}l@{\hspace{2\tabcolsep}\extracolsep{\fill}}r r r r@{}}
\toprule
& \multicolumn{4}{c}{BD-rate (\%) $\downarrow$} \\
\cmidrule(lr){2-5}
Method & CER & WER & \shortstack{ROI\\Similarity}
& \multicolumn{1}{c@{}}{\textbf{Average}} \\
\midrule
\rowcolor{tablegray}
\multicolumn{5}{c}{\textit{VAE-based methods}} \\
MS-ILLM (anchor) & 0.00 & 0.00 & 0.00 & 0.00 \\
GLC     & 85.98 & 215.54 & 11.93 & 104.48 \\
RDVQ    & 28.98 & 42.43 & 11.05 & 27.49 \\
\midrule
\rowcolor{tablegray}
\multicolumn{5}{c}{\textit{Diffusion-based methods}} \\
RDEIC (2-step)   & -3.74  & 6.08   & 17.32  & 6.55 \\
OSCAR            & 210.29 & 298.42 & 137.66 & 215.46 \\
OneDC            & \underline{-30.08} & -10.15 & \underline{-13.94} & \underline{-18.06} \\
StableCodec      & 0.95   & -0.76  & -12.75 & -4.19 \\
CoD              & 831.45 & 582.44 & 136.07 & 516.65 \\
DiT-IC           & 55.74  & 38.85  & -2.96  & 30.54 \\
\midrule
\rowcolor{tablegray}
\multicolumn{5}{c}{\textit{Gen2-IC variants}} \\
Gen2-IC-SD
& 4.16 & 17.53 & -4.89 & 5.60 \\
Gen2-IC-SANA
& -4.58 & 16.14 & -6.82 & 1.58 \\
Gen2-IC-FLUX
& -25.24 & \underline{-18.00} & -1.90 & -15.04 \\
Gen2-IC-Qwen
& \textbf{-36.53} & \textbf{-29.70} & \textbf{-27.01} & \textbf{-31.08} \\
\bottomrule
\end{tabular*}
\endgroup
\caption{Recognizer-based semantic-fidelity comparison.
BD-rate (\%; lower is better) relative to MS-ILLM for CER, WER, and face-ROI
similarity over four benchmarks.}
\label{tab:semantic_bd_rate_ms_illm}
\end{table}

\section{Experiments}
\subsection{Experimental Setup}
\paragraph{Training.}
We instantiate Gen2-IC with SD-2.1~\cite{stable-diffusion},
SANA-1.5~\cite{sana1.5}, FLUX.1-dev~\cite{flux2024}, and
Qwen-Image-2512~\cite{qwenimagetechnicalreport}, using training and validation
subsets from ImageNet~\cite{imagenet} and
OpenImages~\cite{kuznetsova2020open}. 
Each variant uses its native VAE, latent codec, scheduler conversion, and LoRA adapter; the two-stage formulation is shared. 
Latent Compression and Latent Transport use $256\times256$ and $512\times512$ crops, while the codec supports five rates indexed by $\mathrm{qp}\in\{0,\ldots,4\}$. For transport, the codec, VAE, and backbone are frozen; one LoRA adapter per model
($r=16$, $\alpha=32$, dropout $0.0$) is trained at $\mathrm{qp}=0$ and reused
across rates. The specific $\beta_{\mathrm{rate}}$ values are selected on the validation
set and fixed during testing, without per-image or per-dataset adjustment
(e.g., $\{0.99,0.85,0.55,0.45,0.30\}$ for Gen2-IC-Qwen). Model-specific hyperparameter selection is provided in the appendix.

\paragraph{Evaluation.}
We evaluate on Kodak~\cite{kodak}, Tecnick~\cite{tecnick}, DIV2K~\cite{div2k}, and CLIC 2020 Professional~\cite{clic2020}. 
We report BD-rate~\cite{bd-rate} (PCHIP over the common metric range without extrapolation; negative indicates savings), PSNR, MS-SSIM~\cite{ms-ssim}, LPIPS~\cite{lpips} with VGG features~\cite{vgg}, DISTS~\cite{dists}, MUSIQ~\cite{MUSIQ}, and CLIP-IQA~\cite{CLIP-IQA}. FID~\cite{fid} and KID~\cite{kid} follow the GLC protocol~\cite{vq-gic}. 
For recognizer-based semantic evaluation, CRAFT~\cite{CRAFT} and
SCRFD~\cite{SCRFD} detect source text and face ROIs, respectively, which are
reused across all reconstructions (348 text ROIs in 86 images; 215 face ROIs
in 74 images). CER/WER~\cite{Levenshtein} measure normalized edit distances
between source and reconstructed EasyOCR transcripts using CRNN~\cite{CRNN},
while face-ROI similarity is the cosine similarity of their ArcFace
embeddings~\cite{ArcFace}. These source-referenced proxies complement, not replace, dedicated ground-truth OCR and identity-verification evaluations.

\paragraph{Baselines.}
We compare against representative baselines: VTM-23.10~\cite{vtm} and DCVC-RT~\cite{dcvc-rt} in intra mode; VAE-based codecs MS-ILLM~\cite{ms-illm}, GLC~\cite{vq-gic}, and RDVQ~\cite{RDVQ}; and diffusion-based codecs RDEIC~\cite{rdeic}, ResULIC~\cite{ResULIC}, OSCAR~\cite{oscar}, StableCodec~\cite{StableCodec}, OneDC~\cite{oneDC}, CoD~\cite{CoD}, and DiT-IC~\cite{DiT-IC}.

\subsection{Main Results}
\paragraph{Rate-Quality Performance.}
Figure~\ref{fig:rate_quality_curves} compares representative curves across
distortion, perceptual, distributional, and semantic metrics. The Average
column of Table~\ref{tab:bd_rate_ms_illm} aggregates PSNR, LPIPS, and DISTS
across four datasets: Gen2-IC variants consistently balance fidelity and
perceptual quality, whereas several generative codecs incur larger PSNR
penalties for LPIPS/DISTS gains; Gen2-IC-Qwen ranks best at $-32.22\%$.
These gains are not scale-only: using the same SD-2.1 prior as OSCAR,
Gen2-IC-SD achieves a lower BD-rate in every reported fidelity, perceptual,
and semantic metric. Table~\ref{tab:semantic_bd_rate_ms_illm} further shows
the recognizer-based semantic average improving monotonically across SD, SANA,
FLUX, and Qwen, from $5.60\%$ to $1.58\%$, $-15.04\%$, and $-31.08\%$.
Together with the fidelity/perceptual results, this trend shows that Latent
Transport enables stronger priors to improve source-referenced text and face
consistency across Gen2-IC variants, rather than perceptual quality through
source drift.

\paragraph{Qualitative Comparisons.}
Figure~\ref{fig:Visual_example_0} compares challenging text and portrait
regions with recent generative codecs, while Figure~\ref{fig:visual_example_1}
broadens the comparison across representative codec families and scene content.
At comparable or lower bitrates, Gen2-IC-Qwen more faithfully preserves
source-specific text, facial features, and local geometry; competing
reconstructions more often exhibit blur, deformation, or semantic drift.
Additional results are provided in the appendix.

\begin{table}[t]
\centering
\begingroup
\small
\setlength{\tabcolsep}{0.6mm}
\begin{tabular*}{\linewidth}{@{}l@{\extracolsep{\fill}}rrrr@{}}
\toprule
& \multicolumn{4}{c}{BD-rate (\%) $\downarrow$} \\
\cmidrule(lr){2-5}
Model variant & PSNR & MS-SSIM & LPIPS & DISTS \\
\midrule
FLUX: VAE only          & -7.35  & -0.54  & 87.27  & 136.52 \\
FLUX: VAE + Prior       & -12.50 & -9.21  & 19.13  & 33.08  \\
Qwen: VAE only          & 15.79  & 13.77  & 57.90  & 78.32  \\
Qwen: VAE + Prior       & -12.93 & -6.79  & -4.74  & 7.14   \\
\bottomrule
\end{tabular*}
\endgroup
\caption{Generative-prior contribution on Kodak.
BD-rate (\%; lower is better) relative to the pixel-codec baseline used by the
FLUX and Qwen variants.}
\label{tab:kodak_components}
\end{table}

\begin{table}[t]
\centering
\begingroup
\small
\setlength{\tabcolsep}{0pt}
\begin{tabular*}{\linewidth}{@{}l@{\hspace{0.8mm}\extracolsep{\fill}}c@{\hspace{1mm}}r
                    @{\hspace{1mm}}r@{\hspace{1mm}}r@{}}
\toprule
Method & $S$
& \shortstack{Enc.\\$C/T$}
& \shortstack{Dec.\\$C/T$}
& \shortstack{Model\\$P/M$} \\
\midrule
\rowcolor{tablegray}
\multicolumn{5}{c}{\textit{Diffusion-based methods}} \\
RDEIC       & 2 & 2.26/157.25 &  7.60/426.68 &  1.38/6.86 \\
ResULIC     & 3 & 2.08/96.24  &  8.79/428.32 &  1.39/7.99 \\
OSCAR       & 1 & 2.11/65.13  &  6.07/163.77 &  1.01/4.74 \\
OneDC       & 1 & 2.62/100.50 &  6.46/235.03 &  1.41/8.02 \\
StableCodec & 1 & 2.54/105.18 &  6.19/200.03 &  1.07/6.70 \\
CoD         & 1 & 1.65/18.12  &  6.75/62.19  &  0.66/4.21 \\
DiT-IC      & 1 & 2.76/97.62  &  4.54/151.89 &  1.07/5.07 \\
\midrule
\rowcolor{tablegray}
\multicolumn{5}{c}{\textit{Gen2-IC variants}} \\
Gen2-IC-SD   & 1 & 2.32/27.87 &  6.39/103.05  &  1.00/2.70 \\
Gen2-IC-SANA & 1 & 2.51/31.75 &  5.54/102.75  &  1.98/8.19 \\
Gen2-IC-FLUX & 1 & 2.33/21.82 & 38.83/193.58  & 12.17/32.75 \\
Gen2-IC-Qwen & 1 & 3.59/24.05 & 59.96/355.11  & 20.64/56.40 \\
\bottomrule
\end{tabular*}
\endgroup
\caption{Computational complexity on Kodak at $512\!\times\!768$, measured on a single NVIDIA A100 GPU.
$S$: scheduler updates; $C/T$: MMAC/pixel and ms/image; $P/M$: parameters (B)
and peak memory (GiB).}
\label{tab:complexity_kodak}
\end{table}

\begin{table}[t]
\centering
\begingroup
\small
\setlength{\tabcolsep}{1mm}
\begin{tabular*}{\linewidth}{@{}c@{\extracolsep{\fill}}c c c@{}}
\toprule
$t_{\mathrm{inj}}$ & $\beta^\star$ & w/o LoRA & w/ LoRA \\
\midrule
1  & 0.01 & 0.1941 & 0.1847 \\
5  & 0.01 & 0.1919 & 0.1812 \\
10 & 0.05 & 0.1890 & 0.1765 \\
15 & 0.05 & 0.1858 & 0.1748 \\
20 & 0.05 & 0.1827 & 0.1737 \\
25 & 0.05 & 0.1793 & 0.1720 \\
30 & 0.15 & 0.1768 & 0.1711 \\
35 & 0.25 & 0.1743 & 0.1696 \\
40 & 0.45 & 0.1728 & 0.1668 \\
45 & 0.80 & 0.1722 & \underline{0.1577} \\
50 & 0.99 & 0.1716 & \textbf{0.1427} \\
\bottomrule
\end{tabular*}
\endgroup
\caption{Near-terminal injection ablation.
DISTS ($\downarrow$) for Gen2-IC-Qwen at $\mathrm{qp}=0$ on Kodak, using
$\beta^\star$ selected on the validation set for each $t_{\mathrm{inj}}$.
We compare the frozen backbone with its LoRA-adapted counterpart after
100 updates; $t_{\mathrm{inj}}=50$ is the native endpoint of the generation schedule.}
\label{tab:terminal_injection_scan_main}
\end{table}

\paragraph{Complexity Analysis.}
Table~\ref{tab:complexity_kodak} reports full-model inference without VAE tiling or slicing, model sharding, or sparse-attention acceleration. We use four warm-up runs followed by ten timed runs, and include arithmetic coding in the reported latency. Using the same SD-2.1 generative model as OSCAR, Gen2-IC-SD reduces encoding/decoding latency from $65.13/163.77$ to $27.87/103.05$ ms and peak memory from $4.74$ to $2.70$ GiB, while improving all reported quality metrics. Because the generative backbone is used only during decoding, the FLUX and Qwen variants retain fast encoding at $21.82$ and $24.05$ ms, respectively. Adaptation also remains lightweight across model scales: Latent Transport requires only $18.68$M and $23.59$M trainable LoRA parameters, corresponding to $0.16\%$ and $0.12\%$ of the FLUX.1-dev and Qwen-Image-2512 backbones. Additional runtime and memory results are provided in the appendix.

\subsection{Ablation Studies}
We ablate prior components and injection timing. Appendix results
show that activation matters more than LoRA rank and prompts bring no consistent
gain, with visual comparisons.

\paragraph{Generative-Prior Contribution.}
Table~\ref{tab:kodak_components} isolates Latent Transport by holding the VAE
and latent codec fixed. Adding the prior improves all four metrics for FLUX
and Qwen, with matched SD and SANA results in the supplement, demonstrating
effectiveness across distinct latent spaces and generative formulations.
Despite favorable training-aligned metrics, the pixel-codec baseline shows
weaker fidelity and recognizer-based semantic preservation, while supplementary
user-study results show lower human preference.
Using MS-ILLM as the BD-rate anchor, Gen2-IC-Qwen with Latent Transport reduces
the CER, WER, and face-ROI BD-rates from $7.49/-9.33/-15.46\%$ for the
pixel-codec baseline to $-36.53/-29.70/-27.01\%$, respectively.
These results demonstrate that Latent Transport improves
reconstruction quality and recognizer-based semantic preservation.

\paragraph{Near-Terminal Injection.}
We scan Qwen's native 50-step trajectory at $\mathrm{qp}=0$. At each injection
position, $\beta^\star$ is selected on the validation set using the frozen
backbone without LoRA, and then fixed while the LoRA adapter is trained for
100 updates. Table~\ref{tab:terminal_injection_scan_main} reports the resulting
Kodak DISTS. Later injection monotonically improves both variants, from
$0.1941/0.1847$ at $t_{\mathrm{inj}}=1$ to $0.1716/0.1427$ at the native
endpoint $t_{\mathrm{inj}}=50$. The consistent endpoint optimum before and
after adaptation supports near-terminal correction, with LoRA further
amplifying its benefit.

\section{Conclusion}
We introduced Gen2-IC, a simple, extensible interface coupling compression and
generation through separately optimized Latent Compression and one-step Latent
Transport. Using decoded latents as near-terminal states, its direct,
single-path design supports fast encoding and multiple rates without extra
conditioning or task-specific backbone modifications. The shared formulation
applies across SD-2.1, SANA-1.5, FLUX.1-dev, and Qwen-Image-2512: the matched
SD-2.1 comparison isolates interface gains from model scale, while FLUX and
Qwen demonstrate how stronger priors further advance ultra-low-rate
compression.

\FloatBarrier

\clearpage
\appendix
\setcounter{table}{0}
\setcounter{figure}{0}
\setcounter{equation}{0}
\setcounter{algorithm}{0}
\renewcommand{\thetable}{S\arabic{table}}
\renewcommand{\thefigure}{S\arabic{figure}}
\renewcommand{\theequation}{S\arabic{equation}}
\renewcommand{\thealgorithm}{S\arabic{algorithm}}

The main paper presents the core idea and performance of Gen2-IC.  In this
supplementary material, we provide additional experimental details, results,
and analyses, including:
\begin{itemize}
  \item \textbf{Section~\ref{sec:supp_method}: Detailed Method.} Inference,
  native scheduler conversion, codec architecture, and transport objective.
  \item \textbf{Section~\ref{sec:supp_implementation}: Implementation and
  Reproducibility.} Training, validation, evaluation, and baseline settings.
  \item \textbf{Section~\ref{sec:supp_quantitative}: Additional Results.} Matched
  prior contributions, complete quality curves, human preference, and
  multi-resolution efficiency.
  \item \textbf{Section~\ref{sec:supp_ablations}: Additional Ablations.} LoRA
  activation/rank and conditioning without prompt transmission.
  \item \textbf{Section~\ref{sec:supp_visual_results}: Visual Results.}
  Matched-rate source-fidelity comparisons and module-level evidence in
  this appendix.
\end{itemize}

\section{Detailed Method}
\label{sec:supp_method}

\subsection{End-to-End Inference}
\label{sec:supp_algorithm}

\begin{algorithm}[H]
\caption{Gen2-IC inference at operating point $q$}
\label{alg:gen2ic_inference}
\footnotesize
\begin{algorithmic}[1]
\REQUIRE Image $x$; operating point $q$; Gen2-IC modules; native scheduler
$\mathcal S$; fixed $\tau_\star$ and $\beta_q$
\ENSURE Bitstream $\mathcal B$ and reconstruction $\hat x_1$
\STATE $l\leftarrow\mathcal E_e(x)$
\STATE $y\leftarrow g_a(l,q_e^{(q)})$
\STATE $\mathcal B\leftarrow\mathrm{EntropyEncode}(y;\phi)$
\STATE $\hat y\leftarrow\mathrm{EntropyDecode}(\mathcal B;\phi)$
\STATE $\hat l_0\leftarrow g_s(\hat y,q_d^{(q)})$
\STATE $(s_\star,m_\star)\leftarrow
\mathrm{ConstructState}_{\mathcal S}(\hat l_0,\tau_\star)$
\STATE $p_\star\leftarrow P_\theta(s_\star,\tau_\star;m_\star)$
\STATE $s_1\leftarrow\mathrm{Step}_{\mathcal S}(p_\star,\tau_\star,s_\star)$
\STATE $\tilde l_1\leftarrow\mathrm{ToVAELatent}_{\mathcal S}(s_1,m_\star)$
\STATE $\hat l_1\leftarrow\hat l_0+\beta_q(\tilde l_1-\hat l_0)$
\STATE $\hat x_1\leftarrow\mathcal E_d(\hat l_1)$
\STATE \textbf{return} $\mathcal B,\hat x_1$
\end{algorithmic}
\end{algorithm}

Algorithm~\ref{alg:gen2ic_inference} expands the $\Delta_\theta$ interface in
the main paper.  The encoder comprises the frozen VAE encoder and Latent
Compression.  The decoder reconstructs $\hat l_0$, converts it to the last
nontrivial native state, and applies one scheduler update.  Its output
$\tilde l_1$ defines $\Delta_\theta=\tilde l_1-\hat l_0$, scaled by $\beta_q$.
Thus, ``one-step'' denotes exactly one native scheduler update; end-to-end
latency is reported in the main paper and Table~\ref{tab:runtime_resolution}.

The metadata $m_\star$ stores layout, image IDs, sequence shape, prompt masks,
and schedule indices; deterministic seeds generate construction noise.  The
positive prompt is empty, leaving $\hat l_0$ as the image-specific signal
without an image prompt, semantic feature, or auxiliary bitstream.

\subsection{Native Scheduler Conversion}
\label{sec:supp_scheduler}

Table~\ref{tab:native_interface} details SD-2.1~\cite{stable-diffusion},
SANA-1.5~\cite{sana1.5}, FLUX.1-dev~\cite{flux2024}, and
Qwen-Image-2512~\cite{qwenimagetechnicalreport}.  Each uses its last nontrivial
native interval while retaining model-specific state, prediction, scheduler,
adapter, and latent operations.

\paragraph{Latent coordinates.}
SD uses $l=z\,s_{\mathrm{VAE}}$; SANA applies analogous DC-AE scaling.  FLUX
uses $l=(z-s_{\mathrm{shift}})s_{\mathrm{scale}}$, and Qwen uses
$l_c=(z_c-\mu_c)/\sigma_c$.  Both stages share these native coordinates.  SD
and SANA retain BCHW layout; FLUX and Qwen apply
$B\times16\times H_l\times W_l\rightarrow
B\times(H_lW_l/4)\times64$ packing before prediction and invert it afterward;
Qwen also retains its singleton temporal dimension.

\paragraph{Terminal-state construction.}
We use native 20-step SD/SANA and 50-step FLUX/Qwen schedules, executing only
their final update.  FLUX/Qwen retain the sequence-length shift and use sigmas
from $1$ to $1/50$ with \texttt{scale\_noise}.  All variants use seed 42, the
official default negative prompt, and an empty positive prompt; CFG is 5 for
Qwen and 1 otherwise.  Outputs return to VAE coordinates before
$\hat l_1=(1-\beta_q)\hat l_0+\beta_q\tilde l_1$.

\subsection{Latent Codec Architecture}
\label{sec:supp_architecture}

\paragraph{Trainable module.}
The native VAE and backbone remain unchanged.  Latent Compression contains
analysis/synthesis transforms $g_a,g_s$ and entropy model $\phi$, using
depthwise-convolution blocks for parameter-efficient nonlinear capacity.  Its
representation $y$ is split into four quadtree groups whose Gaussian parameters
$(\mu_i,\sigma_i)$ are predicted from decoded groups.  Channel-wise
$q_e^{(q)},q_d^{(q)}$ and entropy-model quantization scales provide five rates
in one codec.

\subsection{What the Transport Objective Learns}
\label{sec:supp_transport_objective}

After freezing Latent Compression, we back-propagate
$\mathcal L_1+\mathcal L_{\mathrm{per}}+\mathcal L_{\mathrm{adv}}$ through the
frozen VAE decoder and near-terminal update to LoRA.  This source-referenced
loss directly trains a residual correction at the decoded latent.  With the
VAE and codec fixed, transport improves PSNR, MS-SSIM, LPIPS, and DISTS.
Source anchoring and terminal placement jointly constrain refinement to remain
faithful while recovering compression-impaired details.

\section{Implementation and Reproducibility}
\label{sec:supp_implementation}

\subsection{Training and Validation Protocol}
\label{sec:supp_training}

We retain the 10,000 highest-resolution images from each of
ImageNet~\cite{imagenet} and OpenImages~\cite{kuznetsova2020open}.  The 30
lowest-resolution images in this combined pool form validation; all others
form training.  Latent Compression samples random
$256\times256$ crops and $q\in\{0,\ldots,4\}$, with rate weight log-linearly
interpolated over $[0.1,1.0]$; Latent Transport uses $512\times512$ crops.
We set the Python, NumPy, and PyTorch training seeds to 42 for every variant.
Transport freezes the codec, VAE, and backbone and optimizes attention LoRA
only.  Each adapter~\cite{lora} uses rank 16, $\alpha=32$, zero dropout, and
targets \texttt{to\_q}, \texttt{to\_k}, \texttt{to\_v}, and
\texttt{to\_out.0}.

Each variant trains one adapter at $\mathrm{qp}=0$ and reuses it across rates.
Validation selects its five $\beta_{\mathrm{rate}}$ values within the codec's
bitrate regime.  Table~\ref{tab:variant_hyperparameters} reports the
variant-specific $\beta_{\mathrm{rate}}$ values, frozen before all four test
sets without per-image or per-dataset adjustment.

The stages are optimized sequentially: Latent Compression is frozen before
LoRA adaptation, separating coding from generative refinement without joint
training.

\begin{table}[htbp]
\centering
\begingroup
\small
\setlength{\tabcolsep}{1.0mm}
\begin{tabular*}{\columnwidth}{@{}l@{\extracolsep{\fill}}l@{}}
\toprule
Variant & $\beta_{\mathrm{rate}}$ for $\mathrm{qp}=0,1,2,3,4$ \\
\midrule
Gen2-IC-SD   & $\{0.95,0.80,0.60,0.45,0.30\}$ \\
Gen2-IC-SANA & $\{0.95,0.80,0.60,0.45,0.30\}$ \\
Gen2-IC-FLUX & $\{0.99,0.95,0.65,0.55,0.40\}$ \\
Gen2-IC-Qwen & $\{0.99,0.85,0.55,0.45,0.30\}$ \\
\bottomrule
\end{tabular*}
\endgroup
\caption{Validation-selected $\beta_{\mathrm{rate}}$ values, fixed before testing.}
\label{tab:variant_hyperparameters}
\end{table}

\subsection{Evaluation Datasets and Metrics}
\label{sec:supp_metrics}

Following prior codecs, we use these standard benchmarks to align evaluation
sets and enable direct baseline comparison.

\begin{table}[htbp]
\centering
\begingroup
\footnotesize
\centering
\textbf{(a) Test datasets}\par\smallskip
\setlength{\tabcolsep}{1.2mm}
\begin{tabular*}{\columnwidth}{@{}l@{\extracolsep{\fill}}rr@{}}
\toprule
Dataset & Images & Resolution \\
\midrule
Kodak~\cite{kodak}              & 24  & $768\times512$ \\
Tecnick~\cite{tecnick}          & 100 & $1200\times1200$ \\
DIV2K val.~\cite{div2k}         & 100 & $\sim$2K \\
CLIC 2020 Pro.~\cite{clic2020}  & 250 & up to 2K \\
\bottomrule
\end{tabular*}
\par\medskip
\centering
\textbf{(b) Reported metrics}\par\smallskip
\setlength{\tabcolsep}{1.2mm}
\begin{tabular}{@{}p{0.27\columnwidth}p{0.57\columnwidth}c@{}}
\toprule
Category & Metric & Better \\
\midrule
Rate             & bpp                                      & $\downarrow$ \\
Fidelity         & PSNR; MS-SSIM                             & $\uparrow$ \\
Perceptual       & LPIPS-VGG; DISTS                          & $\downarrow$ \\
Distributional   & FID; KID                                  & $\downarrow$ \\
No-reference     & MUSIQ; CLIP-IQA                           & $\uparrow$ \\
Text consistency & CER; WER                                  & $\downarrow$ \\
Face consistency & face-ROI cosine similarity               & $\uparrow$ \\
Curve summary    & BD-rate                                   & $\downarrow$ \\
\bottomrule
\end{tabular}
\endgroup
\caption{Evaluation datasets and metrics; FID/KID follow GLC and omit Kodak.}
\label{tab:evaluation_overview}
\end{table}

Each $(\text{variant},\text{dataset},\mathrm{qp})$ uses one deterministic run
with seed 42.  RGB inputs are
replicate-padded to multiples of 32 (SD/FLUX/Qwen) or 128 (SANA), coded, and
cropped before scoring.

\paragraph{Rate and image metrics.}
Rate is $8N_{\mathrm{IP}}/(HW)$, where $N_{\mathrm{IP}}$ is the emitted
intra-picture stream size; the reusable sequence header is excluded.  PSNR is
$-10\log_{10}\mathrm{MSE}$ on RGB $[0,1]$ tensors.  Saved images are scored by
\texttt{torchmetrics} MS-SSIM~\cite{ms-ssim} (\texttt{data\_range=1}) and
LPIPS-VGG~\cite{lpips} with $[0,1]\!\rightarrow\![-1,1]$ normalization.
DISTS~\cite{dists}, MUSIQ~\cite{MUSIQ}, and CLIP-IQA~\cite{CLIP-IQA} use
\texttt{pyiqa}.  The latter two score reconstructions only; all per-image
scores are averaged by dataset and operating point.

\paragraph{Distributional metrics.}
FID/KID~\cite{fid,kid} follow GLC~\cite{vq-gic}.  We extract non-overlapping
$256\times256$ patches on regular and $(128,128)$-shifted grids, quantize to
uint8, and update \texttt{torchmetrics} Inception-2048 statistics.  KID uses
100 subsets of size $\min(50,N_{\mathrm{patch}})$.  Each dataset/rate is scored
over its full patch distribution.  FID/KID use Tecnick, DIV2K, and CLIC 2020,
whose sizes support patch-distribution estimates; Kodak contains only 24
images and is not used for these metrics.

\paragraph{BD-rate.}
BD-rate~\cite{bd-rate} uses PCHIP interpolation of log-rate over each pair's
common metric interval, without extrapolation.  Samples are metric-sorted;
duplicate metric values are merged by mean rate, and all remaining overlapping
samples are retained.  Comparisons require a nonempty common interval.
Negative values denote bitrate savings at matched quality.

\paragraph{Recognizer-based ROI construction.}
EasyOCR uses CRAFT~\cite{CRAFT} with Chinese/English recognition.  Source text
boxes require confidence $\geq0.3$ and at least two alphanumeric/CJK
characters.  InsightFace \texttt{buffalo\_l} uses SCRFD~\cite{SCRFD} at
$640\times640$ and threshold 0.5.  Detectors run once on sources, and fixed
boxes are reused across reconstructions.  Table~\ref{tab:semantic_roi_counts}
reports 348 text boxes in 86 images and 215 face boxes in 74 images; 56 and 24
images contain multiple boxes, with per-image maxima of 27 and 54.

\begin{table}[htbp]
\centering
\begingroup
\footnotesize
\textbf{(a) Text ROIs}\par\smallskip
\setlength{\tabcolsep}{1.2mm}
\begin{tabular*}{\columnwidth}{@{}l@{\extracolsep{\fill}}rrrr@{}}
\toprule
Dataset & Images & ROIs & Multi & Max/image \\
\midrule
Kodak     & 2  & 5   & 2  & 3  \\
Tecnick   & 10 & 24  & 6  & 5  \\
DIV2K     & 16 & 64  & 13 & 11 \\
CLIC 2020 & 58 & 255 & 35 & 27 \\
\midrule
Total     & 86 & 348 & 56 & 27 \\
\bottomrule
\end{tabular*}
\par\medskip
\textbf{(b) Face ROIs}\par\smallskip
\begin{tabular*}{\columnwidth}{@{}l@{\extracolsep{\fill}}rrrr@{}}
\toprule
Dataset & Images & ROIs & Multi & Max/image \\
\midrule
Kodak     & 7  & 16  & 3  & 8  \\
Tecnick   & 1  & 1   & 0  & 1  \\
DIV2K     & 17 & 57  & 7  & 19 \\
CLIC 2020 & 49 & 141 & 14 & 54 \\
\midrule
Total     & 74 & 215 & 24 & 54 \\
\bottomrule
\end{tabular*}
\endgroup
\caption{Source ROIs.  Multi counts multi-region images; Max/image is the
largest per-image count.}
\label{tab:semantic_roi_counts}
\end{table}

\paragraph{Text consistency.}
Each source text box is expanded by 5\% and mapped to the reconstruction.
EasyOCR/CRNN~\cite{CRNN} outputs with confidence $\geq0.3$ are ordered by
reading order, lowercased, and restricted to alphanumeric/CJK characters.  CER
removes spaces; WER tokenizes them.  Levenshtein distance is source-length
normalized.  Scores are averaged first by image, then by dataset.

\paragraph{Face consistency.}
Source face boxes and five-point landmarks are mapped to reconstructions.
ArcFace~\cite{ArcFace} embeds aligned $112\times112$ crops; when direct
alignment is unavailable, detection is repeated within a 20\%-enlarged fixed
box.  Cosine similarity of
$\ell_2$-normalized embeddings is averaged by image, then dataset.  The
evaluation includes boxes down to $11\times16$ pixels; minimum/median SCRFD scores are
0.501/0.783.  CER/WER use source EasyOCR outputs, while face consistency uses
source/reconstruction ArcFace embeddings within the fixed ROIs.

\subsection{Baseline and Complexity Settings}
\label{sec:supp_baselines}

\paragraph{Classical codec.}
VTM-23.10~\cite{vtm} uses one-frame all-intra coding with 8-bit YUV 4:2:0 input
and the official configuration in Table~\ref{tab:vtm_configuration}.

\begin{table}[htbp]
\centering
\begingroup
\footnotesize
\setlength{\tabcolsep}{1.4mm}
\begin{tabular*}{\columnwidth}{@{}l@{\extracolsep{\fill}}l@{}}
\toprule
Setting & Value \\
\midrule
Executable & \texttt{EncoderApp} \\
Configuration & \texttt{encoder\_intra\_vtm.cfg} \\
Input / reconstruction & \texttt{input.yuv} / \texttt{output.yuv} \\
Bitstream & \texttt{output.bin} \\
Resolution & \texttt{--wdt [width] --hgt [height]} \\
Quality point & \texttt{-q [QP]} \\
Bit depth / chroma & 8-bit / YUV 4:2:0 \\
Frames / frame rate & 1 / 1 \\
\bottomrule
\end{tabular*}
\endgroup
\caption{VTM-23.10 all-intra evaluation configuration.}
\label{tab:vtm_configuration}
\end{table}

\paragraph{Neural codecs and complexity.}
DCVC-RT also runs in intra mode; neural-codec rates use actual entropy-coded
streams.  \texttt{calflops} measures MACs with $1$ FLOP $=2$ MACs.  A100
latency includes arithmetic coding and averages ten runs after four warm-ups.
We disable VAE tiling/slicing, sharding, sparse attention, and
deployment-specific backbone optimization.

\section{Additional Quantitative Results}
\label{sec:supp_results}
\label{sec:supp_quantitative}

\subsection{Matched Generative-Prior Contribution}
\label{sec:supp_prior_contribution}

\paragraph{Controlled comparison.}
Table~\ref{tab:matched_sd_sana} extends the main paper's matched FLUX/Qwen
ablation to SD/SANA.  Each pair fixes the native VAE and codec, changing only
Latent Transport, which improves every metric.  Across four native latent
spaces and formulations, these matched comparisons establish consistent
within-variant gains from Latent Transport.

\begin{table}[htbp]
\centering
\begingroup
\small
\setlength{\tabcolsep}{0.7mm}
\begin{tabular*}{\linewidth}{@{}l@{\extracolsep{\fill}}rrrr@{}}
\toprule
& \multicolumn{4}{c}{BD-rate (\%) $\downarrow$} \\
\cmidrule(lr){2-5}
Variant & PSNR & MS-SSIM & LPIPS & DISTS \\
\midrule
SD: Latent Compression & 114.52 & 39.50 & -48.88 & -65.14 \\
SD: + Latent Transport & 71.82 & 25.53 & -62.65 & -72.64 \\
SANA: Latent Compression & 144.77 & 54.94 & -59.91 & -73.36 \\
SANA: + Latent Transport & 114.28 & 39.89 & -69.72 & -79.58 \\
\bottomrule
\end{tabular*}
\endgroup
\caption{Matched SD/SANA Kodak ablation relative to MS-ILLM, using the
latent-codec and corresponding one-step LoRA curves with each VAE and codec
fixed.}
\label{tab:matched_sd_sana}
\end{table}

\paragraph{Interpretation.}
These matched gains directly isolate Latent Transport and show consistent
distortion and perceptual improvements across all four native latent spaces.

\subsection{Additional Quality Metrics}
\label{sec:supp_additional_metrics}

\begin{figure*}[p]
\centering
\includegraphics[width=0.94\textwidth]{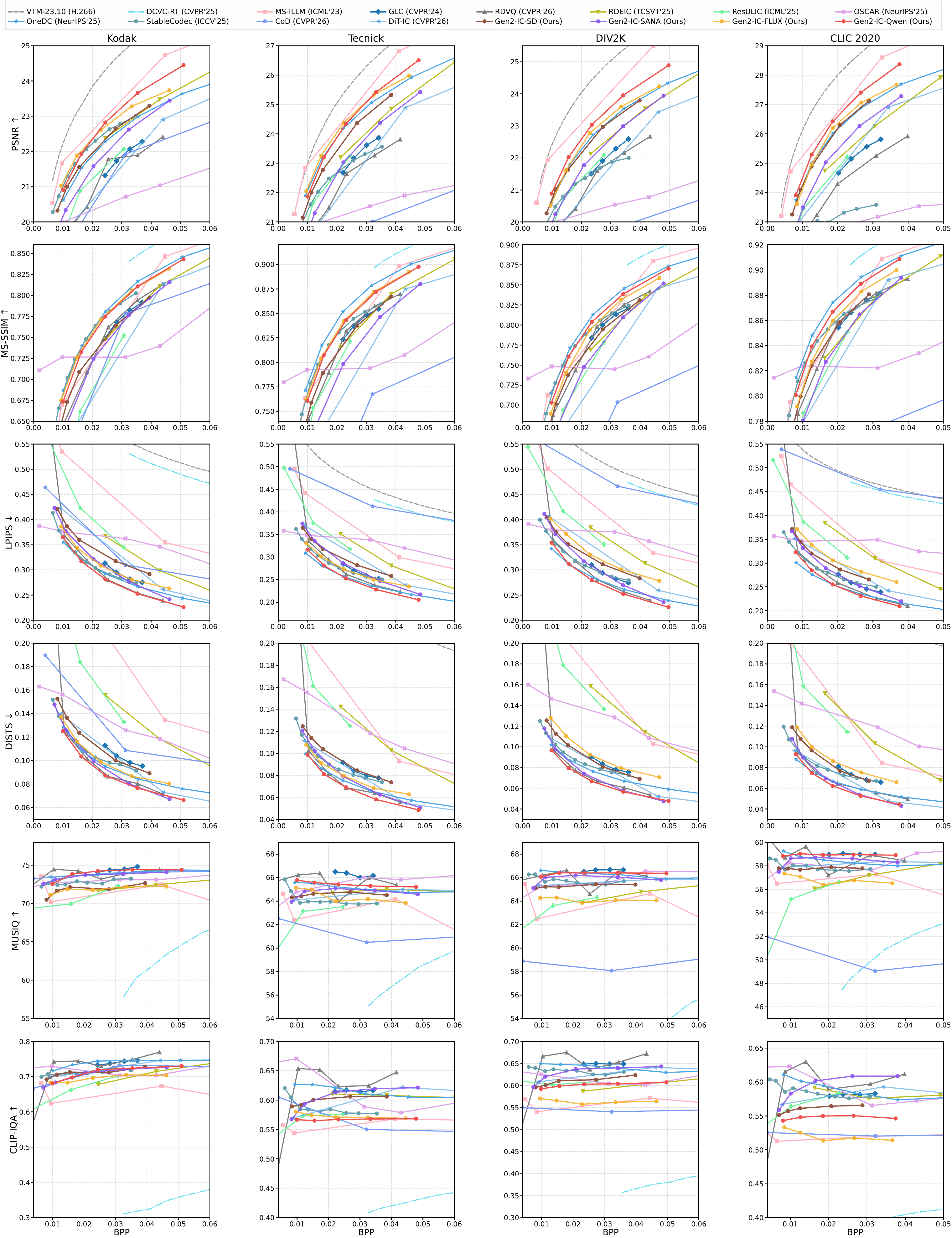}
\caption{PSNR, MS-SSIM, LPIPS, DISTS, MUSIQ, and CLIP-IQA rate--quality
curves on all four benchmarks.}
\label{fig:supp_quality_curves}
\end{figure*}

\begin{figure*}[p]
\centering
\includegraphics[width=0.96\textwidth]{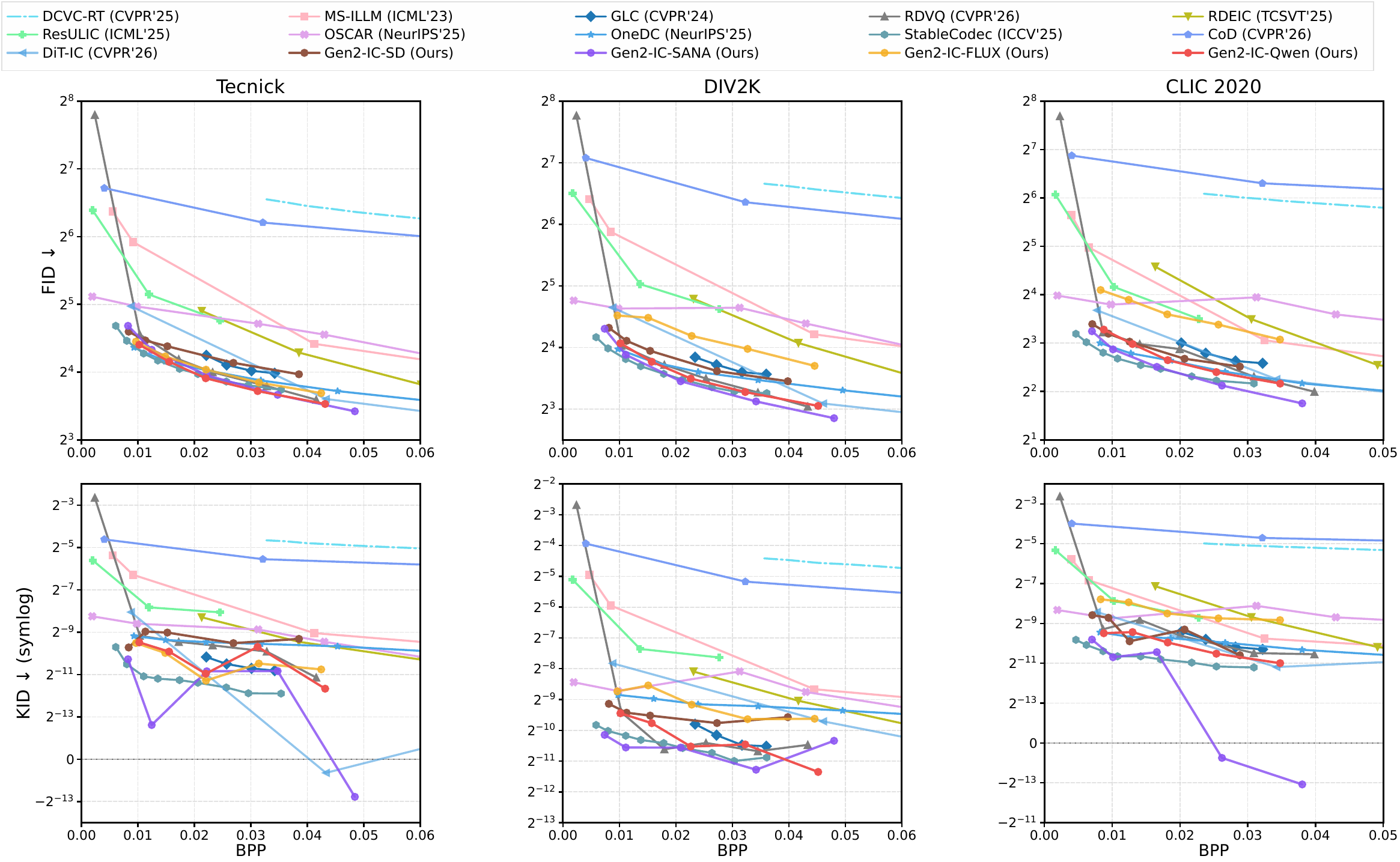}
\caption{FID/KID rate--quality curves on Tecnick, DIV2K, and CLIC 2020.}
\label{fig:supp_distribution_curves}
\end{figure*}

\paragraph{Complementary curves.}
Figures~\ref{fig:supp_quality_curves} and~\ref{fig:supp_distribution_curves}
provide complete PSNR, MS-SSIM, LPIPS, DISTS, MUSIQ~\cite{MUSIQ}, and
CLIP-IQA~\cite{CLIP-IQA} curves on all four benchmarks, together with FID/KID
on Tecnick, DIV2K, and CLIC 2020.  They extend the main results across
distortion, perceptual, no-reference, and distributional criteria.

\subsection{User Study}
\label{sec:supp_user_study}

\paragraph{Protocol.}
Following StableCodec's rate-matched protocol~\cite{StableCodec}, 33
participants select the best reconstruction among randomly ordered methods for
each of 24 Kodak images, yielding $33\times24=792$ top-1 votes.

\paragraph{Result.}
Gen2-IC-Qwen ranks first with 268 votes (33.84\%), 1.65$\times$ the runner-up
and 3.57$\times$ Gen2-IC-Base (pixel codec).  This preference gap complements
the quantitative results and corroborates Gen2-IC's human-perceived
reconstruction quality.

\begin{table}[htbp]
\centering
\begingroup
\small
\setlength{\tabcolsep}{1.5mm}
\begin{tabular*}{\columnwidth}{@{}l@{\extracolsep{\fill}}rr@{}}
\toprule
Method & Top-1 votes & Share (\%) \\
\midrule
\textbf{Gen2-IC-Qwen} & \textbf{268} & \textbf{33.84} \\
OneDC                  & 162 & 20.45 \\
StableCodec            & 100 & 12.63 \\
Gen2-IC-Base (pixel codec) & 75  & 9.47 \\
GLC                    & 71  & 8.96 \\
OSCAR                  & 39  & 4.92 \\
MS-ILLM                & 34  & 4.30 \\
DiffEIC~\cite{diffeic} & 27  & 3.41 \\
ResULIC                & 16  & 2.02 \\
\bottomrule
\end{tabular*}
\endgroup
\caption{Blind top-1 preference on 24 rate-matched Kodak cases; shares sum to
100\%.}
\label{tab:user_study}
\end{table}

\subsection{Runtime and Memory Across Resolutions}
\label{sec:supp_resources}

\paragraph{Memory scaling.}
Table~\ref{tab:stage_memory} reports total peak-reserved and stage-incremental
allocated memory.  Its $512\times768$ maxima match the peak memory in main
Table~\ref{tab:complexity_kodak}.  At $1440\times2560$, Latent Transport adds only 0.23--1.59 GiB
beyond resident allocation across all variants, despite $9.4\times$ more
pixels.  Thus, transport has modest resolution-dependent memory overhead even
for large priors.

\paragraph{Runtime scaling.}
Table~\ref{tab:runtime_resolution} reports end-to-end runtime.  At both
$1080\times1920$ and $1440\times2560$, every Gen2-IC variant encodes faster
than all baselines with reported runtimes, and Gen2-IC-SANA decodes fastest.
At the largest resolution, matched Gen2-IC-SD is $10.7\times$ faster in
encoding and $4.6\times$ faster in decoding than OSCAR, confirming that the
interface gain scales beyond Kodak resolution.

\section{Additional Ablations}
\label{sec:supp_ablation}
\label{sec:supp_ablations}

\subsection{LoRA Activation and Rank}
\label{sec:supp_lora}

\paragraph{Prior activation versus rank.}
Table~\ref{tab:lora_rank} separates prior invocation from adaptation in
Gen2-IC-Qwen.  LoRA produces substantial gains over both VAE-only and frozen
prior correction.  Rank 16, $\alpha=32$ performs best overall, while ranks 8
and 32 retain most gains.  For Gen2-IC-Qwen, this confirms that lightweight
prior activation matters more than the exact rank.

\subsection{Prompt Conditioning}
\label{sec:supp_prompts}

\paragraph{Protocol and overhead.}
We compare empty, generic, and compression-oriented prompts.
Qwen3-VL-30B-A3B-Instruct~\cite{Qwen3-VL} generates the latter two at
temperature 0 ($\leq4096$ tokens); lossless \texttt{.xz} overhead averages
0.0002/0.002 bpp for generic/compression prompts.

\begin{figure*}[p]
\centering
\includegraphics[width=0.949\textwidth]{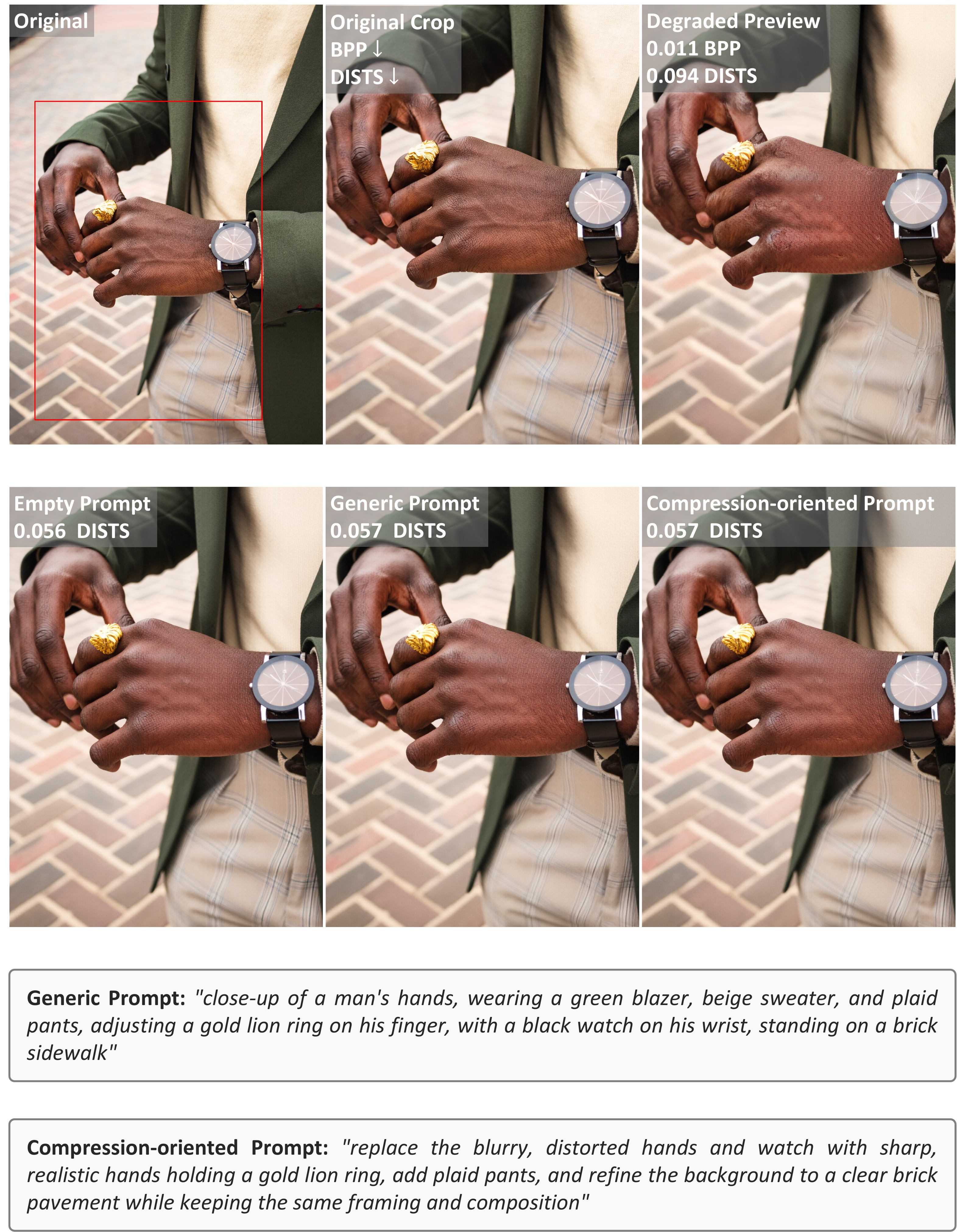}
\caption{Prompt-conditioned comparisons; the empty prompt preserves source
content without transmitted prompt bits.}
\label{fig:supp_prompt_example_0}
\end{figure*}

\begin{figure*}[p]
\centering
\includegraphics[width=0.95\textwidth]{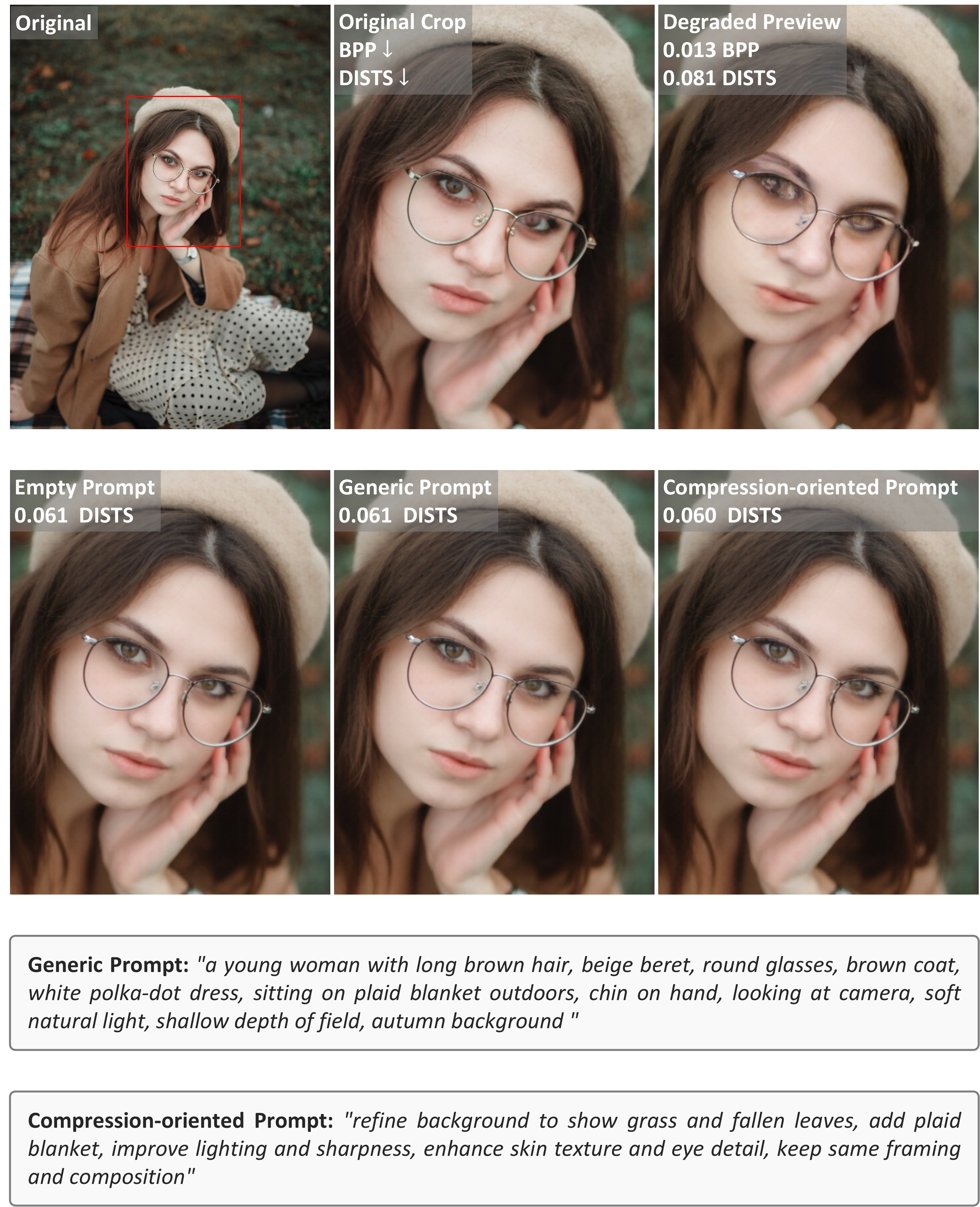}
\caption{Prompt-conditioned comparisons; decoded-latent guidance remains
effective without transmitted prompt bits.}
\label{fig:supp_prompt_example_1}
\end{figure*}

\paragraph{Result.}
For Gen2-IC-Qwen, the empty prompt achieves the strongest overall results in
Table~\ref{tab:prompt_ablation} without prompt transmission.  This confirms
that the decoded latent provides sufficient image-specific guidance and
supports Gen2-IC without extra transmitted conditions.  Prompt
Boxes~\ref{box:generic_prompt_template} and
\ref{box:compression_prompt_template} give the exact templates.  Examples are
included at original resolution in Figures~\ref{fig:supp_prompt_example_0}
and~\ref{fig:supp_prompt_example_1}.

\begin{promptbox}[label={box:generic_prompt_template}]{Generic Prompt Template}
\footnotesize
\textbf{Role.} You are a prompt-writer for image reconstruction.

\textbf{Goal.} Produce a compact instruction that preserves all visually
certain details needed to reconstruct the same content, style, and composition.

\textbf{Requirements.}
\begin{itemize}
  \item Cover certain subjects, layout, background, colors, lighting,
  materials, viewpoint, and style.
  \item Prefer factual nouns and attributes over poetic adjectives.
  \item State an obvious style (e.g., photo, 3D, anime, flat icon, or line art)
  in 1--3 tokens and avoid artist lists.
\end{itemize}
\end{promptbox}

\begin{promptbox}[label={box:compression_prompt_template}]{Compression Prompt Template}
\footnotesize
\textbf{Role.} You are a prompt-writer for image-to-image refinement.

\textbf{Input.} Base image $B$ is the coarse reconstruction and target image
$A$ is the source.

\textbf{Goal.} Describe only the changes needed to transform $B$ into $A$
while preserving matched content and composition.

\textbf{Requirements.}
\begin{itemize}
  \item Keep the prompt compact and explicitly preserve framing/composition.
  \item Do not re-describe the full image.
  \item Focus on missing objects, corrected colors, textures, lighting,
  fine details, sharpness, and artifact removal.
\end{itemize}
\end{promptbox}

These ablations use Gen2-IC-Qwen.
Together, the ablations validate lightweight LoRA activation and operation
without transmitted prompts.

\section{Visual Results}
\label{sec:supp_visual_results}

\begin{figure*}[p]
\centering
\includegraphics[width=\textwidth,height=0.88\textheight,keepaspectratio]{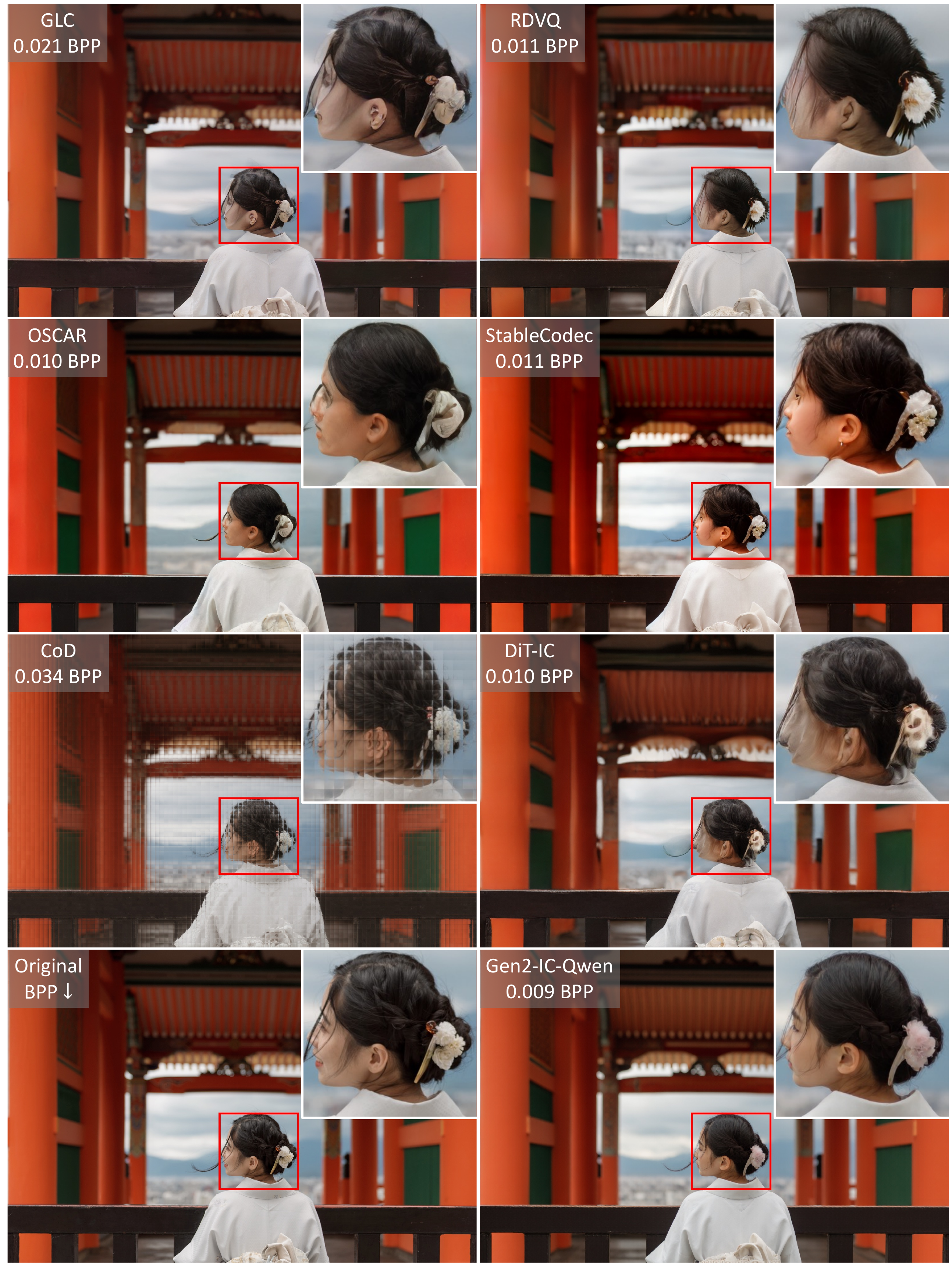}
\caption{Matched-rate comparison on a face-containing scene.  The inset
highlights profile and hair details.}
\label{fig:supp_visual_0}
\end{figure*}

\begin{figure*}[p]
\centering
\includegraphics[width=\textwidth,height=0.88\textheight,keepaspectratio]{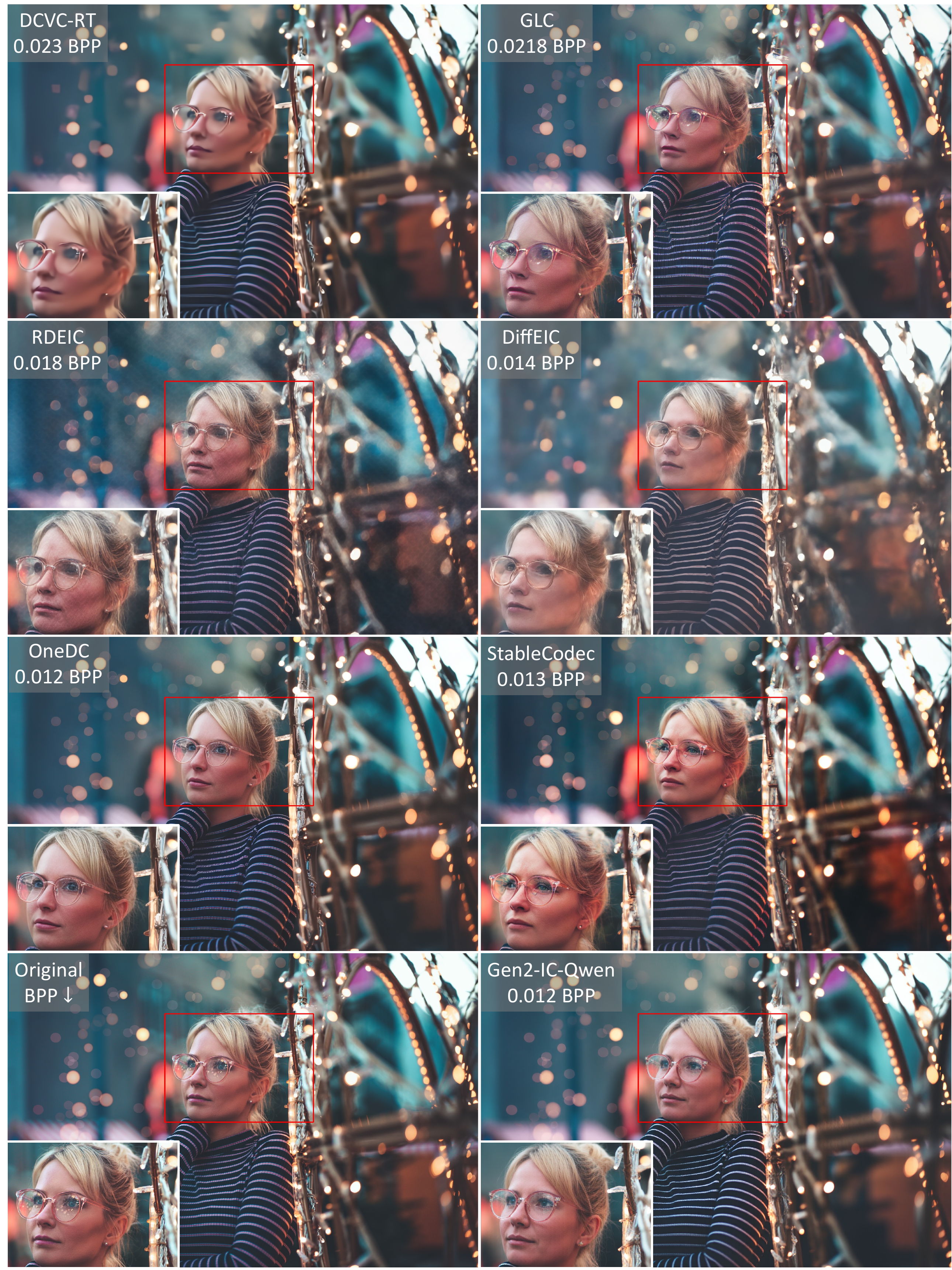}
\caption{Matched-rate comparison on a portrait scene.  The inset highlights
facial details.}
\label{fig:supp_visual_1}
\end{figure*}

\begin{figure*}[p]
\centering
\includegraphics[width=\textwidth,height=0.88\textheight,keepaspectratio]{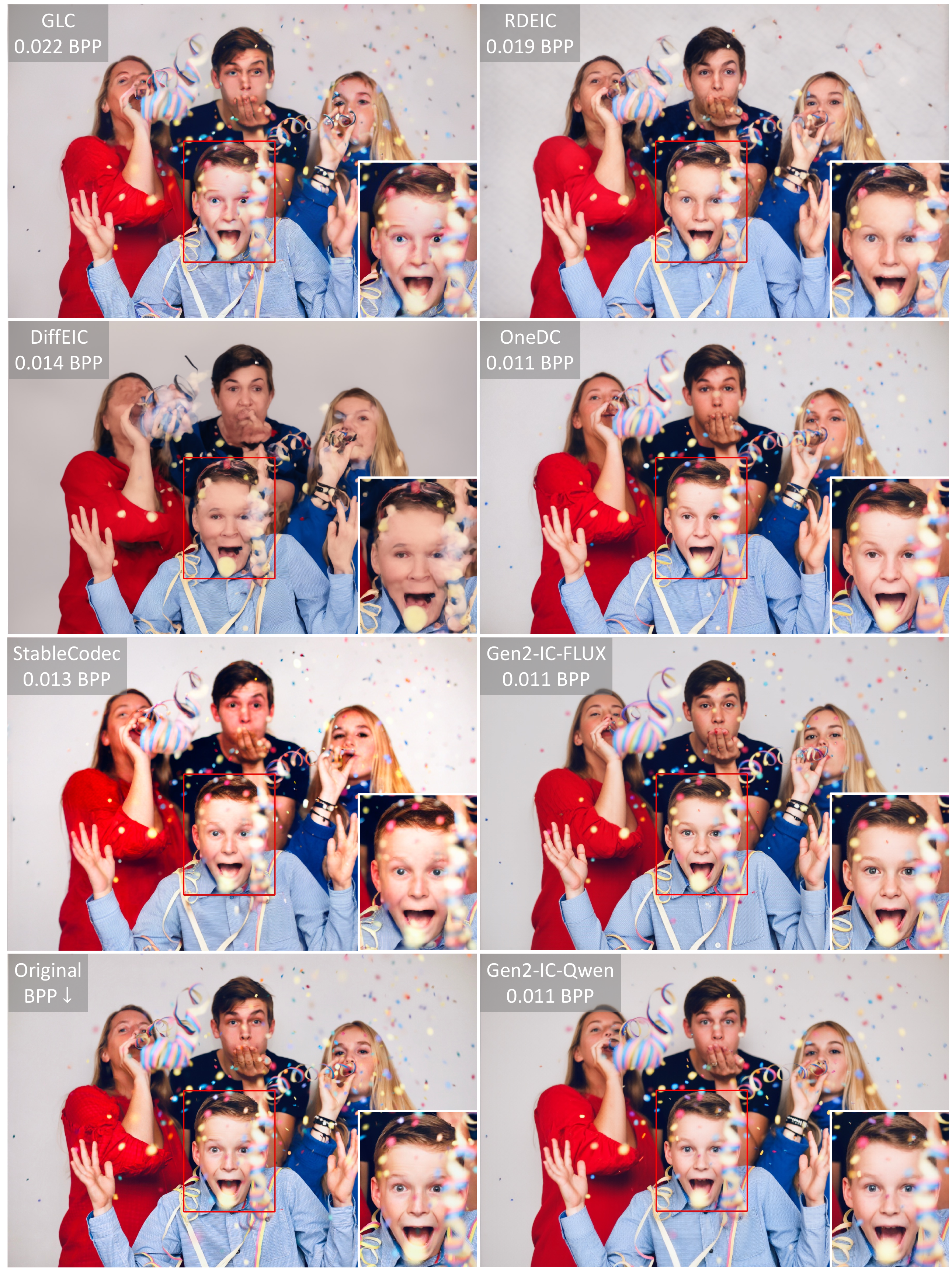}
\caption{Matched-rate comparison on a multi-person scene.  The inset
highlights facial reconstruction.}
\label{fig:supp_visual_2}
\end{figure*}

\begin{figure*}[p]
\centering
\includegraphics[width=\textwidth,height=0.88\textheight,keepaspectratio]{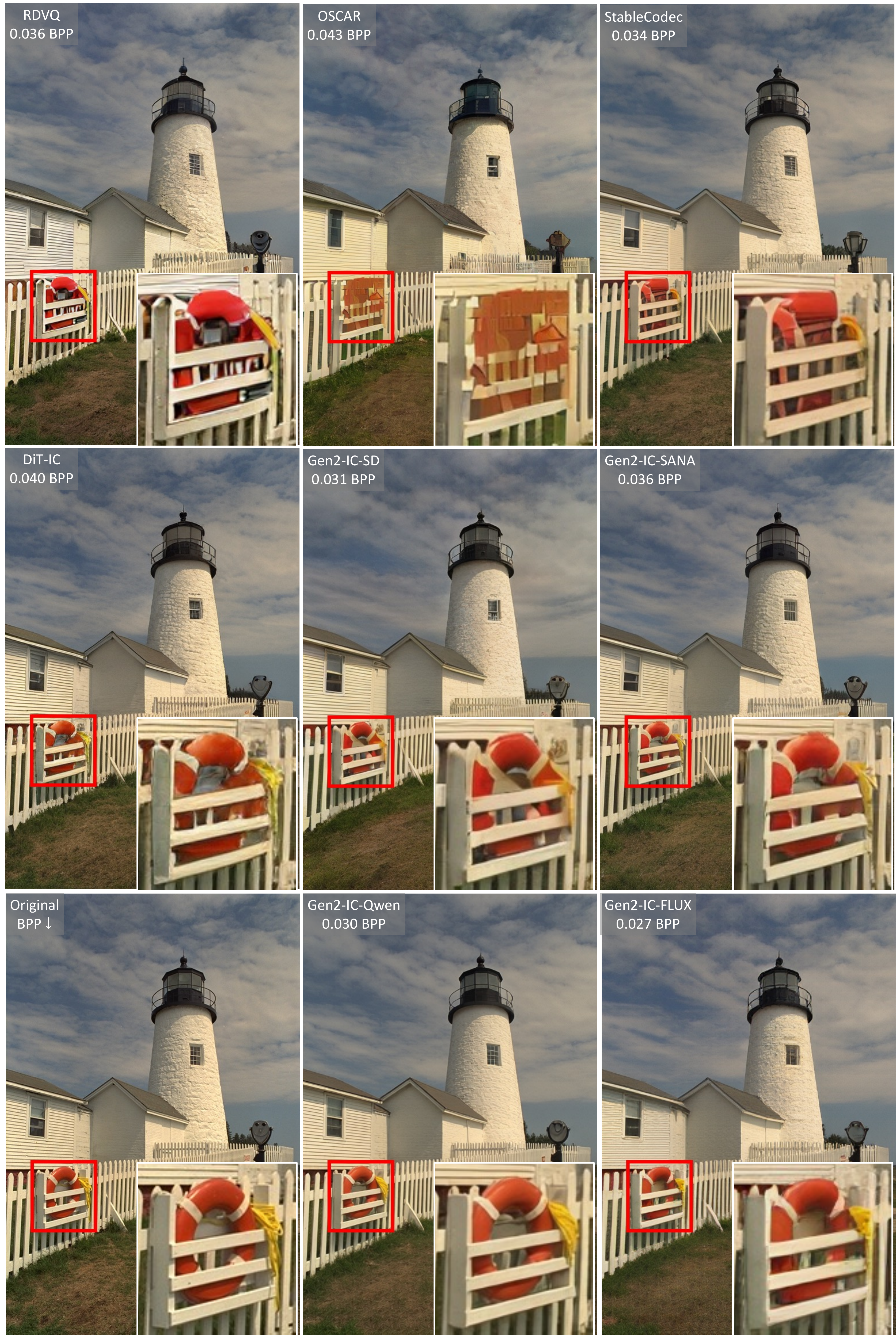}
\caption{Matched-rate comparison on a structured scene.  The inset highlights
the small foreground object.}
\label{fig:supp_visual_3}
\end{figure*}

\begin{figure*}[p]
\centering
\includegraphics[width=\textwidth,height=0.88\textheight,keepaspectratio]{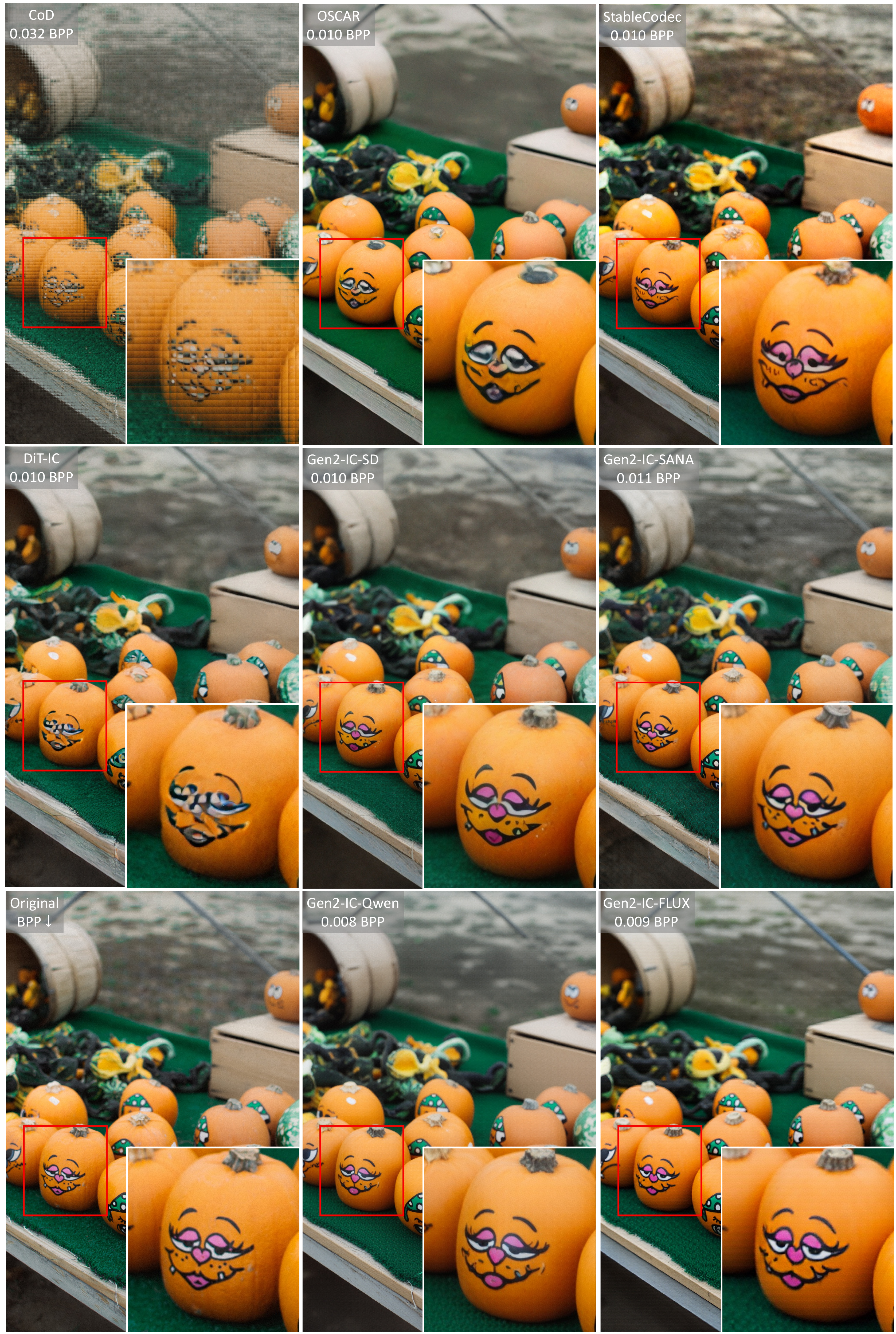}
\caption{Matched-rate comparison on a detail-rich scene.  The inset highlights
the painted pattern.}
\label{fig:supp_visual_4}
\end{figure*}

\begin{figure*}[p]
\centering
\includegraphics[width=\textwidth,height=0.88\textheight,keepaspectratio]{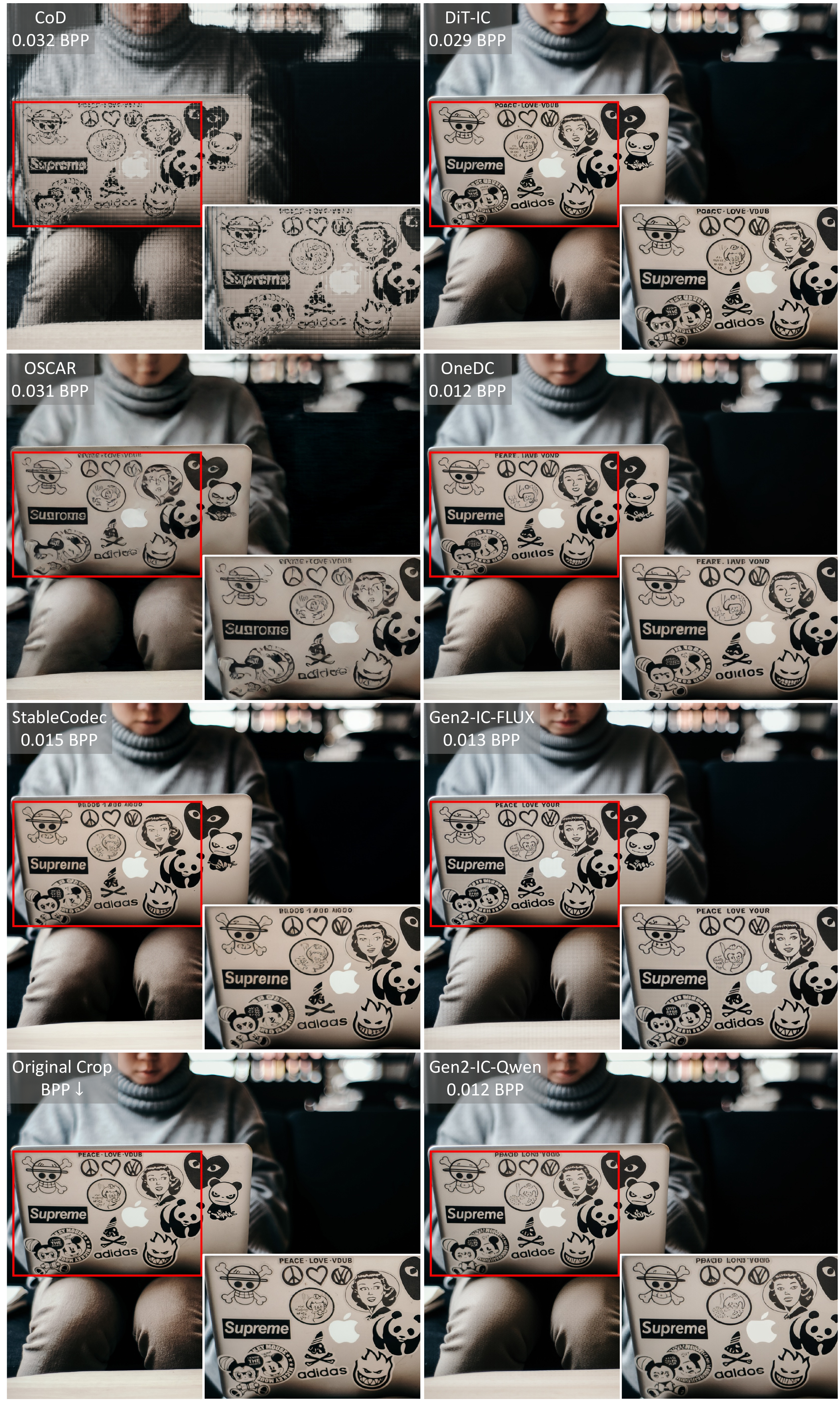}
\caption{Matched-rate comparison on a text- and logo-rich scene.  The inset
highlights fine line structures.}
\label{fig:supp_visual_5}
\end{figure*}

\begin{figure*}[p]
\centering
\includegraphics[width=\textwidth,height=0.88\textheight,keepaspectratio]{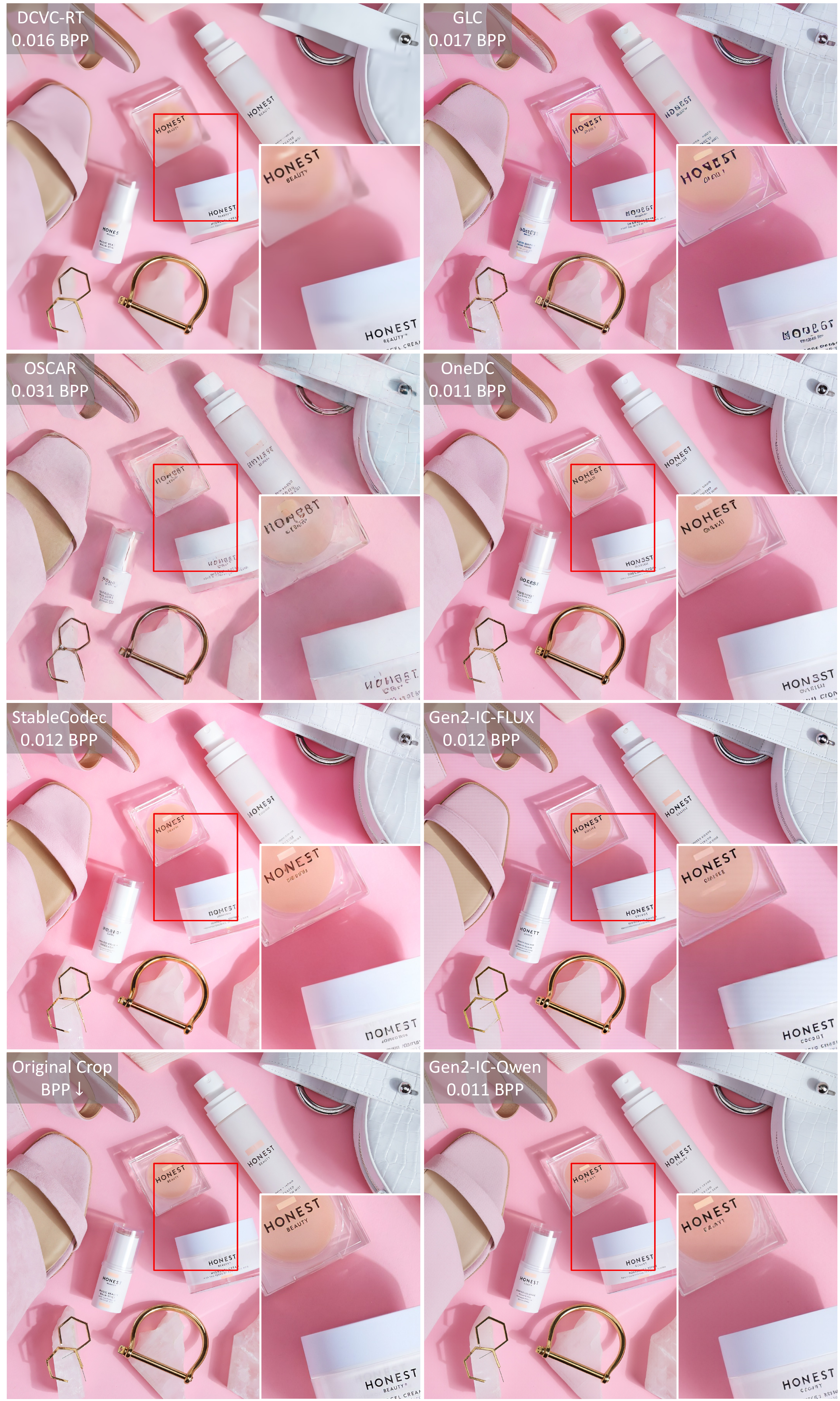}
\caption{Matched-rate comparison on a product scene.  The inset highlights
small printed text.}
\label{fig:supp_visual_6}
\end{figure*}

\begin{figure*}[p]
\centering
\includegraphics[width=\textwidth,height=0.88\textheight,keepaspectratio]{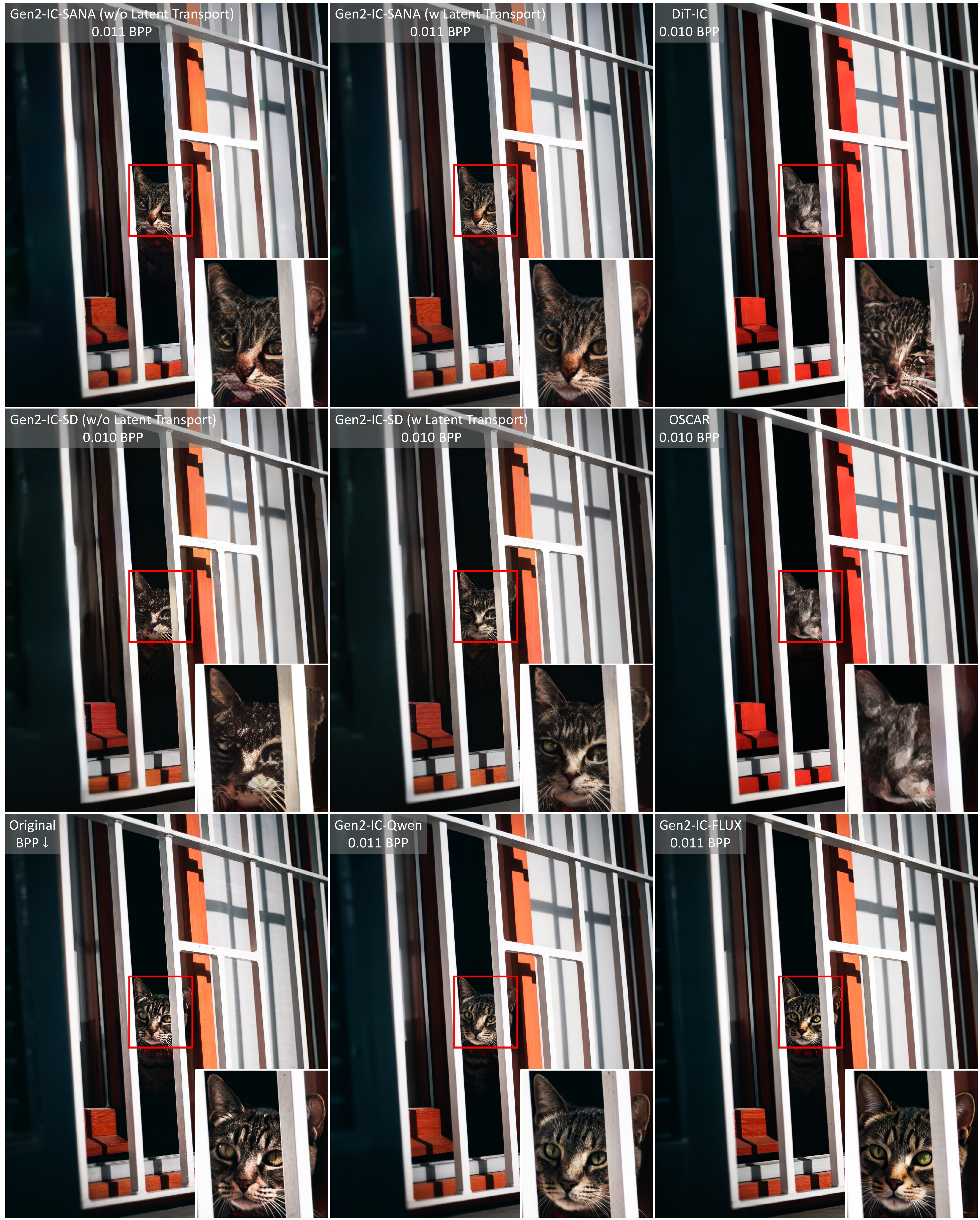}
\caption{Matched-rate component and baseline comparison.  Insets highlight
the effect of Latent Transport on the partially occluded cat.}
\label{fig:supp_visual_ablation_0}
\end{figure*}

\begin{figure*}[p]
\centering
\includegraphics[width=\textwidth,height=0.88\textheight,keepaspectratio]{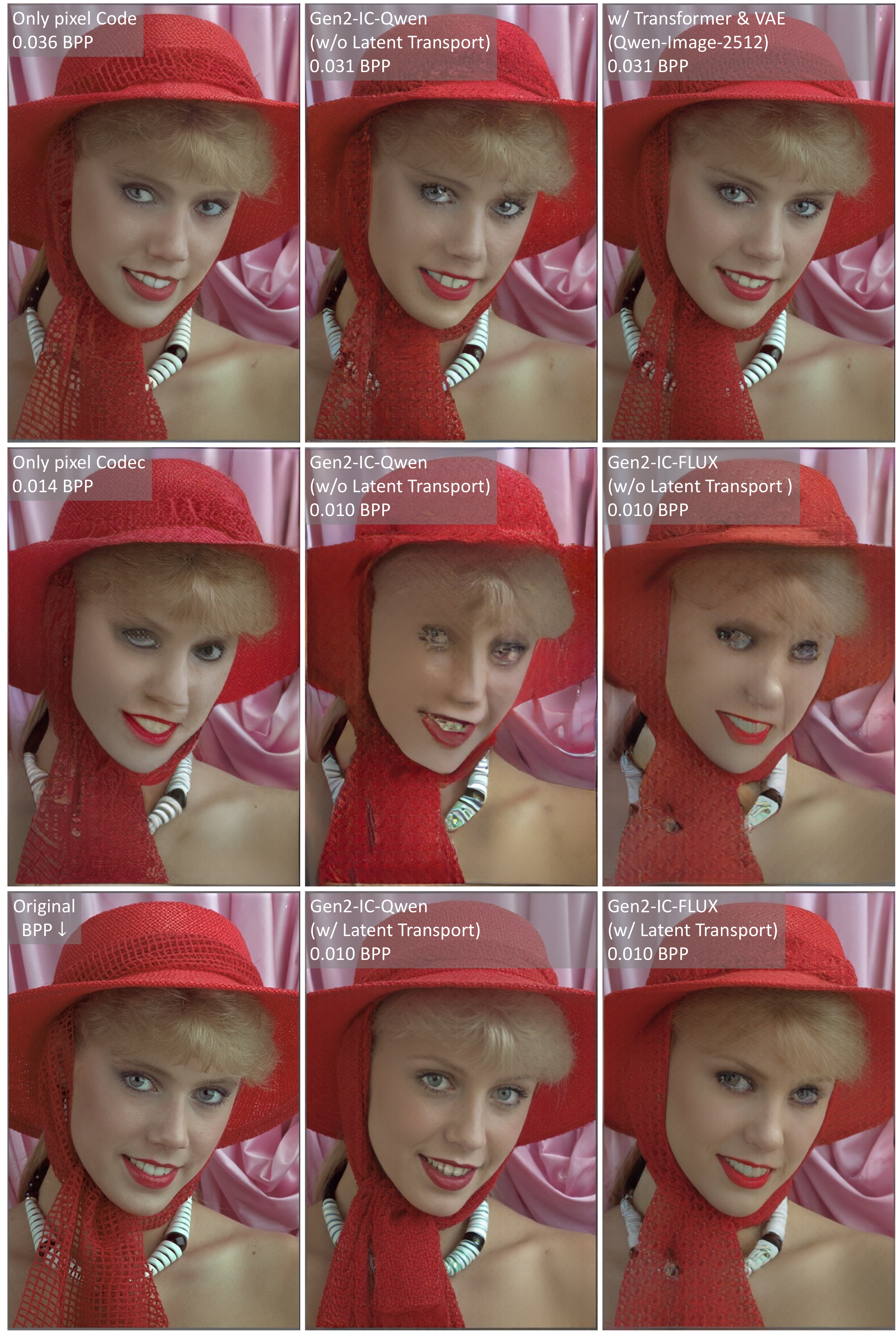}
\caption{Component comparison of Gen2-IC-Base (pixel codec), VAE
reconstruction, and Latent Transport for face reconstruction.}
\label{fig:supp_visual_ablation_1}
\end{figure*}

\begin{figure*}[p]
\centering
\includegraphics[width=\textwidth,height=0.78\textheight,keepaspectratio]{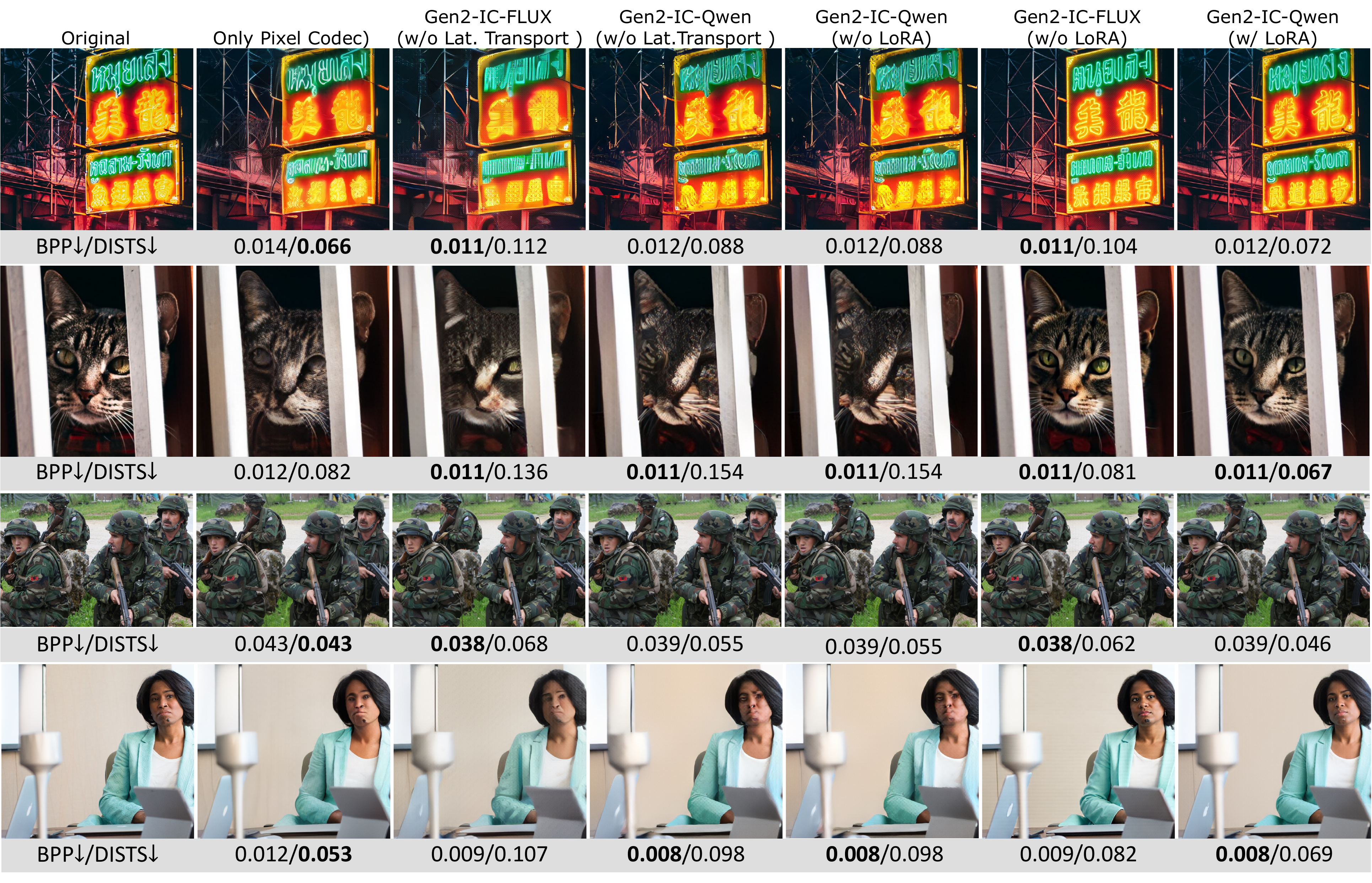}
\caption{Component comparison across Gen2-IC-Base (pixel codec), VAE-only
reconstruction, frozen-prior correction, and LoRA-adapted Latent Transport.
Each row reports bpp/DISTS.}
\label{fig:supp_visual_ablation_2}
\end{figure*}

Figures~\ref{fig:supp_visual_0}--\ref{fig:supp_visual_6} compare matched-rate
reconstructions on face-, structure-, and text-sensitive examples, with actual
bpp and magnified regions. Figures~\ref{fig:supp_visual_ablation_0}--
\ref{fig:supp_visual_ablation_2} isolate Gen2-IC-Base (pixel codec), VAE
reconstruction, frozen-prior correction, and LoRA-adapted Latent Transport,
corroborating the matched quantitative ablations. The original-resolution panels
are embedded in this appendix.

\newpage
\begin{table*}[t]
\centering
\begingroup
\small
\setlength{\tabcolsep}{1.4mm}
\begin{tabular}{@{}lp{0.27\textwidth}p{0.57\textwidth}@{}}
\toprule
Variant & VAE/codec state & State construction and native update \\
\midrule
Gen2-IC-SD & Scaled BCHW latent, $4\times H/8\times W/8$ &
Add native noise at the penultimate state; scale the model input; predict with
the SD-2.1 U-Net; take the final scheduler step \\
Gen2-IC-SANA & Scaled BCHW latent, $32\times H/32\times W/32$ &
Construct the terminal noisy state using native noise/sigma conversion; predict
with the SANA Transformer; take the native step \\
Gen2-IC-FLUX & Shifted/scaled $16$-channel latent; $2\times2$ pack to $64$-D tokens &
Use the resolution-shifted flow schedule, image IDs, and empty-prompt embeddings;
take one native step; unpack to BCHW \\
Gen2-IC-Qwen & Channel-standardized $16$-channel latent; $2\times2$ pack to $64$-D tokens &
Use the resolution-shifted flow schedule, image shapes, and prompt masks; apply
true CFG $=5$; take one native step; unpack to BCHW \\
\bottomrule
\end{tabular}
\endgroup
\caption{Model-specific Latent Transport interfaces; each executes one final
native scheduler update.}
\label{tab:native_interface}
\end{table*}

\begin{table*}[t]
\centering
\begingroup
\small
\setlength{\tabcolsep}{1.8mm}
\begin{tabular*}{\textwidth}{@{}ll@{\extracolsep{\fill}}ccc@{}}
\toprule
Variant & Stage & $512\times768$ & $1080\times1920$ & $1440\times2560$ \\
\midrule
\multirow{5}{*}{Gen2-IC-SD}
& VAE encoding & 1.02 / 0.56 & 4.56 / 3.00 & 7.79 / 5.29 \\
& Latent codec encoding & 0.49 / 0.07 & 1.07 / 0.38 & 1.69 / 0.67 \\
& Latent codec decoding & 0.46 / 0.06 & 0.86 / 0.34 & 1.28 / 0.59 \\
& VAE decoding & 1.45 / 0.89 & 5.97 / 4.72 & 10.32 / 8.33 \\
& Latent Transport & 2.70 / 0.10 & 3.18 / 0.51 & 3.75 / 0.90 \\
\midrule
\multirow{5}{*}{Gen2-IC-SANA}
& VAE encoding & 1.53 / 0.66 & 5.05 / 3.70 & 8.38 / 6.58 \\
& Latent codec encoding & 0.78 / 0.01 & 0.78 / 0.05 & 0.88 / 0.09 \\
& Latent codec decoding & 0.78 / 0.01 & 0.78 / 0.05 & 0.87 / 0.08 \\
& VAE decoding & 1.58 / 0.61 & 5.54 / 3.42 & 9.29 / 6.08 \\
& Latent Transport & 8.19 / 0.03 & 8.38 / 0.13 & 8.53 / 0.23 \\
\midrule
\multirow{5}{*}{Gen2-IC-FLUX}
& VAE encoding & 1.13 / 0.56 & 4.64 / 3.00 & 8.38 / 5.29 \\
& Latent codec encoding & 0.40 / 0.04 & 0.66 / 0.19 & 0.92 / 0.34 \\
& Latent codec decoding & 0.38 / 0.03 & 0.56 / 0.17 & 0.90 / 0.30 \\
& VAE decoding & 1.41 / 0.89 & 5.89 / 4.73 & 10.63 / 8.35 \\
& Latent Transport & 32.75 / 0.21 & 38.77 / 0.91 & 40.14 / 1.56 \\
\midrule
\multirow{5}{*}{Gen2-IC-Qwen}
& VAE encoding & 1.51 / 0.99 & 7.47 / 5.26 & 12.08 / 9.28 \\
& Latent codec encoding & 0.44 / 0.04 & 0.72 / 0.19 & 1.04 / 0.34 \\
& Latent codec decoding & 0.43 / 0.03 & 0.62 / 0.17 & 0.84 / 0.30 \\
& VAE decoding & 2.21 / 1.55 & 9.86 / 8.23 & 17.04 / 14.52 \\
& Latent Transport & 56.40 / 0.17 & 56.64 / 0.90 & 56.98 / 1.59 \\
\bottomrule
\end{tabular*}
\endgroup
\caption{GPU memory across resolutions (GiB); entries report total peak
reserved / stage-incremental allocated memory.}
\label{tab:stage_memory}
\end{table*}

\begin{table*}[t]
\centering
\begingroup
\small
\setlength{\tabcolsep}{0.55mm}
\begin{tabular*}{\textwidth}{@{}l@{\extracolsep{\fill}}*{6}{r}@{}}
\toprule
\multirow{2}{*}{Method}
& \multicolumn{2}{c}{$512\times768$}
& \multicolumn{2}{c}{$1080\times1920$}
& \multicolumn{2}{c}{$1440\times2560$} \\
\cmidrule(lr){2-3}\cmidrule(lr){4-5}\cmidrule(lr){6-7}
& Enc. & Dec. & Enc. & Dec. & Enc. & Dec. \\
\midrule
\rowcolor{tablegray}
\multicolumn{7}{c}{\textit{VAE-based methods}} \\
RDVQ~\cite{RDVQ} & 408.24 & 412.44 & 704.64 & 809.99 & 953.80 & 1130.17 \\
\midrule
\rowcolor{tablegray}
\multicolumn{7}{c}{\textit{Diffusion-based methods}} \\
DiffEIC~\cite{diffeic} & 210.18 & 4661.74 & -- & $>10$ s & -- & $>15$ s \\
RDEIC~\cite{rdeic} & 157.25 & 426.68 & 683.20 & 2296.58 & 1436.00 & 5963.77 \\
ResULIC~\cite{ResULIC} & 96.24 & 428.32 & 481.75 & 3702.45 & 985.24 & 10028.48 \\
OSCAR~\cite{oscar} & 65.13 & 163.77 & 857.68 & 1109.37 & 2982.82 & 2751.47 \\
OneDC~\cite{oneDC} & 100.50 & 235.03 & 366.32 & 763.99 & 680.47 & 1452.33 \\
StableCodec~\cite{StableCodec} & 105.18 & 200.03 & 641.76 & 1657.43 & 1085.39 & 3161.54 \\
CoD~\cite{CoD} & 18.12 & 62.19 & 166.72 & 439.11 & 635.46 & 1866.48 \\
DiT-IC~\cite{DiT-IC} & 97.62 & 151.89 & 480.22 & 750.38 & 718.17 & 1111.59 \\
\midrule
\rowcolor{tablegray}
\multicolumn{7}{c}{\textit{Gen2-IC variants}} \\
Gen2-IC-SD & 27.87 & 103.05 & 144.79 & 323.19 & 279.83 & 600.14 \\
Gen2-IC-SANA & 31.75 & 102.75 & 132.58 & 234.72 & 213.82 & 373.40 \\
Gen2-IC-FLUX & 21.82 & 193.58 & 133.80 & 923.57 & 264.12 & 1807.57 \\
Gen2-IC-Qwen & 24.05 & 355.11 & 130.99 & 2189.06 & 247.65 & 4956.31 \\
\bottomrule
\end{tabular*}
\endgroup
\caption{Runtime (ms/image). Enc./Dec. cover the full coding/reconstruction
pipelines under the same A100 protocol.}
\label{tab:runtime_resolution}
\end{table*}

\begin{table*}[t]
\centering
\begingroup
\small
\setlength{\tabcolsep}{0.65mm}
\begin{tabular*}{\textwidth}{@{}l@{\extracolsep{\fill}}*{12}{r}@{}}
\toprule
\multirow{2}{*}{Qwen variant}
& \multicolumn{4}{c}{PSNR BD-rate $\downarrow$}
& \multicolumn{4}{c}{LPIPS BD-rate $\downarrow$}
& \multicolumn{4}{c}{DISTS BD-rate $\downarrow$} \\
\cmidrule(lr){2-5}\cmidrule(lr){6-9}\cmidrule(lr){10-13}
& Kod. & Tec. & DIV2K & CLIC & Kod. & Tec. & DIV2K & CLIC & Kod. & Tec. & DIV2K & CLIC \\
\midrule
VAE only & 95.35 & 95.11 & 120.19 & 92.67 & -56.57 & -44.31 & -52.16 & -49.12 & -65.68 & -49.34 & -62.61 & -47.73 \\
Transport, frozen prior & 87.88 & 87.40 & 111.71 & 84.54 & -57.42 & -46.04 & -53.37 & -50.69 & -65.16 & -49.83 & -62.72 & -48.13 \\
LoRA $r=8,\alpha=16$ & 48.77 & 49.91 & 73.98 & 53.64 & -74.00 & -65.91 & -68.94 & -66.22 & -79.51 & -65.94 & -75.18 & -65.77 \\
LoRA $r=16,\alpha=32$ & 44.66 & 44.23 & 67.41 & 46.92 & -75.84 & -69.25 & -72.21 & -70.48 & -81.21 & -71.18 & -79.00 & -70.66 \\
LoRA $r=32,\alpha=64$ & 45.89 & 46.86 & 70.04 & 50.08 & -75.03 & -66.55 & -68.58 & -66.62 & -80.32 & -66.10 & -73.66 & -64.99 \\
\bottomrule
\end{tabular*}
\endgroup
\caption{Gen2-IC-Qwen LoRA activation/rank ablation.  BD-rate (\%) is relative
to MS-ILLM.}
\label{tab:lora_rank}
\end{table*}

\begin{table*}[t]
\centering
\begingroup
\small
\setlength{\tabcolsep}{0.55mm}
\begin{tabular*}{\textwidth}{@{}l@{\extracolsep{\fill}}*{12}{r}@{}}
\toprule
\multirow{2}{*}{Setting}
& \multicolumn{4}{c}{PSNR BD-rate (\%) $\downarrow$}
& \multicolumn{4}{c}{LPIPS BD-rate (\%) $\downarrow$}
& \multicolumn{4}{c}{DISTS BD-rate (\%) $\downarrow$} \\
\cmidrule(lr){2-5}\cmidrule(lr){6-9}\cmidrule(lr){10-13}
& Kod. & Tec. & DIV2K & CLIC
& Kod. & Tec. & DIV2K & CLIC
& Kod. & Tec. & DIV2K & CLIC \\
\midrule
No transport
& 0 & 0 & 0 & 0 & 0 & 0 & 0 & 0 & 0 & 0 & 0 & 0 \\
\midrule
Frozen + Empty
& -3.89 & -3.92 & -3.93 & -3.92 & -1.69 & -2.57 & -2.05 & -2.51 & 1.06 & -0.97 & -0.35 & -0.79 \\
Frozen + Generic
& 7.43 & 0.90 & 0.10 & 0.49 & 8.28 & 1.82 & 0.58 & 1.42 & 8.15 & 1.80 & 0.98 & 1.40 \\
Frozen + Compression
& 9.66 & 1.06 & 0.08 & 0.49 & 10.78 & 2.00 & 0.56 & 1.44 & 10.89 & 2.01 & 1.05 & 1.54 \\
\midrule
LoRA + Empty
& -25.37 & -24.44 & -22.02 & -21.04 & -41.89 & -43.11 & -40.08 & -37.73 & -43.13 & -42.22 & -43.49 & -40.13 \\
LoRA + Generic
& -14.19 & -22.15 & -20.97 & -19.21 & -23.75 & -40.29 & -39.83 & -35.39 & -19.77 & -40.01 & -45.95 & -38.92 \\
LoRA + Compression
& -8.82 & -19.77 & -20.09 & -18.49 & -13.87 & -35.12 & -36.68 & -32.48 & -7.89 & -29.62 & -38.15 & -32.75 \\
\bottomrule
\end{tabular*}
\endgroup
\caption{Gen2-IC-Qwen prompt/LoRA ablation.  BD-rate (\%) is relative to no
transport and includes prompt bits.}
\label{tab:prompt_ablation}
\end{table*}

\begin{figure*}[p]
\centering
\includegraphics[width=\textwidth]{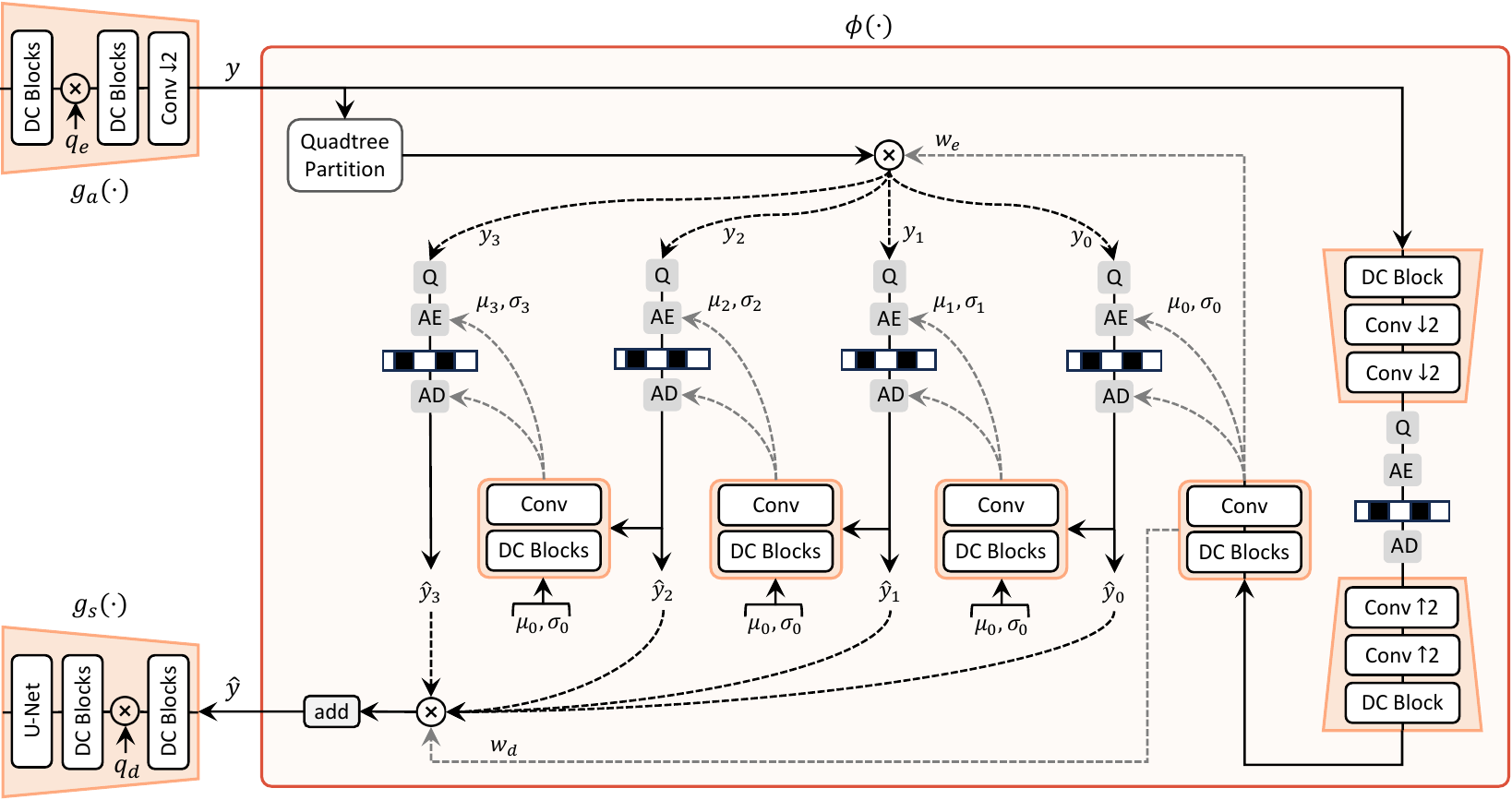}
\caption{Latent Compression: analysis/synthesis transforms and a four-group
quadtree entropy model with arithmetic encoding (AE) and decoding (AD).}
\label{fig:supp_codec_overview}
\end{figure*}

\begin{figure*}[p]
\centering
\includegraphics[width=\textwidth]{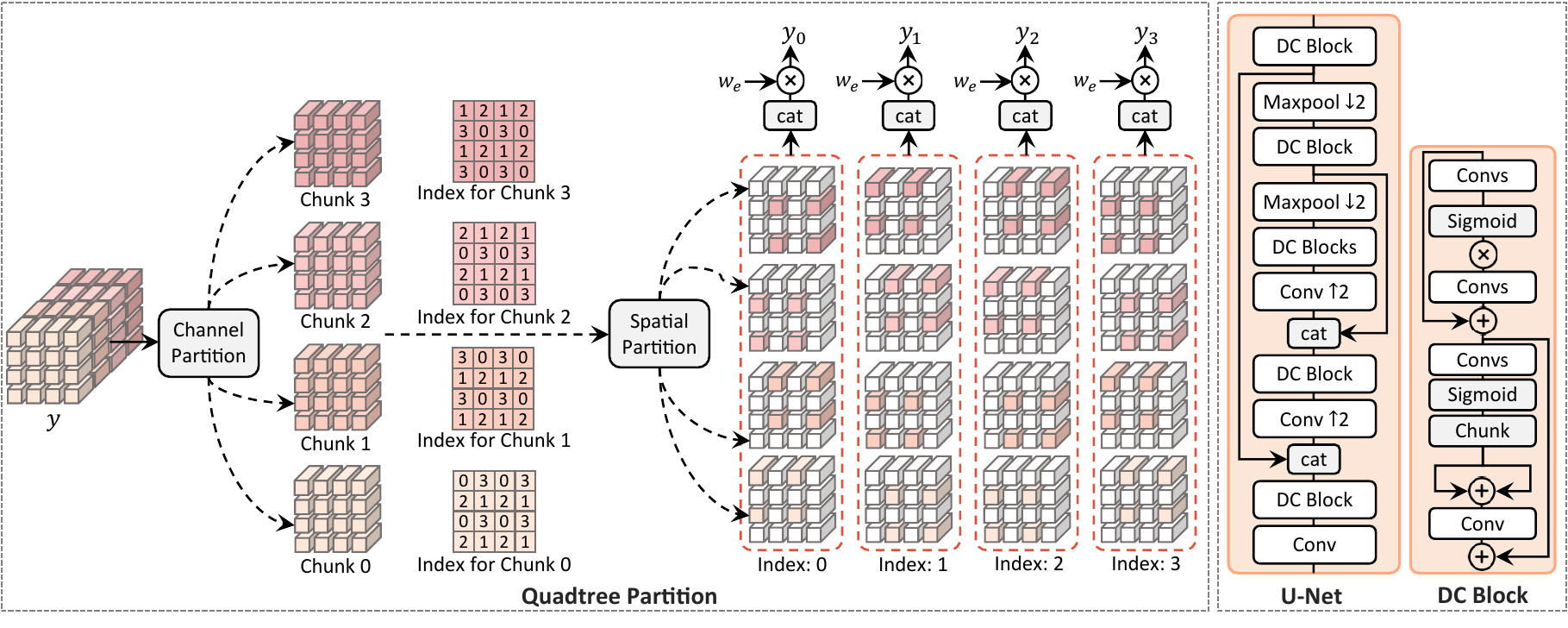}
\caption{Detailed quadtree partition, U-Net transform, and depthwise
convolution (DC) blocks used by Latent Compression.}
\label{fig:supp_codec_details}
\end{figure*}

\clearpage

\end{document}